\RequirePackage{etex}

\documentclass[aps,pra, 10pt,twocolumn,superscriptaddress,nofootinbib,floatfix]{revtex4-2}

\usepackage[english]{babel}

\usepackage[utf8]{inputenc}
\usepackage{amsmath,amssymb,amsfonts,amsthm}
\usepackage{mathtools}
\usepackage{braket}
\usepackage{graphicx}
\usepackage{float}
\usepackage{tikz-network}
\usepackage{tikz}
\usepackage{caption}
\usepackage{subcaption}
\usepackage{algorithm}
\usepackage{algpseudocode}
\usepackage{float}
\usepackage[version=4]{mhchem}
\usepackage[colorlinks=true, bookmarks=false, allcolors=blue]{hyperref}
\usepackage{orcidlink}
\usepackage{lipsum}
\usepackage{hyperref}
\usepackage{xcolor}  

\usepackage{changes}
\algrenewcommand\algorithmicdo{}

\usetikzlibrary{decorations.pathreplacing, decorations.text}

\DeclareUnicodeCharacter{00F6}{\"o}

\DeclarePairedDelimiter\abs{\lvert}{\rvert}
\DeclarePairedDelimiter\norm{\lVert}{\rVert}

\graphicspath{{figures/}}
\makeatletter
\def\input@path{{figures/}}
\makeatother

\begin{document}

\title{An Accurate Lanczos Method for the Matrix Product State Representation}

\author{Yu Wang\,\orcidlink{0009-0004-3972-4388}\textsuperscript{\S}}
\email{18yu.wang@tum.de}
\affiliation{Technical University of Munich, CIT, Department of Computer Science, Boltzmannstra{\ss}e 3, 85748 Garching, Germany}

\author{Zhangyu Yang\,\orcidlink{0009-0009-1589-7985}\textsuperscript{\S}}
\email{zhangyu.yang@tum.de}
\affiliation{Technical University of Munich, NAT, Department of Physics, James-Franck-Straße 1, 85748 Garching, Germany}

\author{Xingyao Wu\,\orcidlink{0009-0009-7762-9038}}
\affiliation{Technical University of Munich, NAT, Department of Physics, James-Franck-Straße 1, 85748 Garching, Germany}

\author{Christian B.~Mendl\,\orcidlink{0000-0002-6386-0230}}
\affiliation{Technical University of Munich, CIT, Department of Computer Science, Boltzmannstra{\ss}e 3, 85748 Garching, Germany}
\affiliation{Technical University of Munich, Institute for Advanced Study, Lichtenbergstra{\ss}e 2a, 85748 Garching, Germany}

\date{\today}

\begin{abstract}
The Lanczos algorithm naturally finds multiple low-lying eigenstates simultaneously, providing a global and robust treatment of the spectrum without local minima or root flipping. However, when implemented within the matrix product state (MPS) formalism, its accuracy is severely affected by MPS truncation, leading to a pronounced high-error plateau. In this work, we propose the modified thick-block Lanczos (MTBL) method to enhance the convergence of the Lanczos algorithm within the MPS formalism. Numerical benchmarks on the Fermi–Hubbard and Heisenberg models demonstrate that MTBL achieves the optimal accuracy allowed by the bond dimension. In addition, we benchmark MTBL on a 50-site Fermi-Hubbard model and a 120-site Heisenberg model to demonstrate its scalability. Furthermore, when incorporating the shift-and-invert technique, MTBL enables the direct targeting of multiple excited states without finding lower eigenstates.

\end{abstract}

\maketitle

\def\thefootnote{{\S}}\footnotetext{These authors contributed equally to this work.}\def\thefootnote{\arabic{footnote}}

\section{Introduction}
Eigenvalue problems lie at the heart of quantum many-body physics, materials science, and quantum chemistry. By finding a few low-lying eigenstates of the Hamiltonian, one can access ground-state properties~\cite{lanczos1950iteration, PhysRevLett.69.2863, SandvikSSE1997, banuls2023routemap, avella2013stronglycorrelated}, low-temperature thermodynamics~\cite{avella2013stronglycorrelated, SugiuraShimizu2013, JaklicPrelovsek2000, laguta2023low}, and dynamical correlation functions~\cite{GaglianoBalseiro1987, SunMotta2021}. The Lanczos algorithm~\cite{lanczos1950iteration, Paige1972, Cullum1985, bai2000templates, Bhattacharya2023, WeinbergBukov2019, LauchliSudanMoessner2019, IskakovDanilov2018} is widely used to find the low-lying eigenstates of large linear systems, and is implemented in popular packages like ARPACK~\cite{LehoucqSorensenYang1998}. Nevertheless, its capability is limited by the curse of dimensionality for many-body quantum systems, as the Hilbert space dimension grows exponentially with the system size. 

Tensor network methods ~\cite{schollwock2011density, Bridgeman_2017, PhysRevLett.75.3537, PhysRevB.55.2164, ma2022density, chan2016matrix, keller2015efficient} representing quantum states and operators by matrix product state (MPS) and matrix product operators (MPO) significantly alleviate this issue. As a key insight, the low-lying states only stay in a small corner of the whole Hilbert space due to the area law of the entanglement entropy, allowing them to be efficiently approximated by low-rank decompositions like MPS~\cite{prgjndfm}. The density-matrix renormalization-group (DMRG) algorithm exploits these properties and optimizes an MPS ansatz variationally. However, DMRG is inherently a local energy minimization method and can become trapped in local minima~\cite{white2005density, pfeifer2015symmetry,dektor2025inexact}. When targeting excited states, it may suffer from root flipping~\cite{baiardi2019optimization, dorando2007targeted, tran2019tracking}, where an unintended excited state is tracked instead of the desired one, especially in the presence of (near-)degeneracies. Moreover, its accuracy depends on the quality of previously obtained lower-energy states, which leads to cumulative errors~\cite{szenes2025qcmaquis4, larsson2019ttns}. The Lanczos algorithm, by contrast, naturally produces a spectrum of eigenpairs and inherently avoids local minima when initialized with the proper initial state. Moreover, through the construction of an orthogonal Krylov subspace, it can inherently resolve degenerate or nearly degenerate low-energy eigenstates when equipped with block initial vectors~\cite{golub1977block, wu1999thickrestart}. Following the pioneering attempts~\cite{dargel2012lanczos, huang2018generalized}, the combination of the Lanczos algorithm and tensor network offers a promising complementary approach for exploring the low-lying eigenstates of large-scale quantum many-body systems~\cite{dektor2025inexact, li2024boundstates, paeckel2023bandlanczos, rano2025inexactlanczos}. 

Unfortunately, using tensor network representations for the Lanczos algorithm introduces a new bottleneck: generating new Krylov vectors enlarges the bond dimensions, so MPS compression is required to keep the calculation feasible and efficient. The MPS truncation introduces non-negligible errors, which result in two significant issues. First, this error causes a loss of orthogonality among the Krylov vectors. A post-reorthogonalization technique~\cite{dargel2012lanczos} has been developed to address this issue. Second, because each Krylov vector depends on its predecessors, the truncation error accumulates as the subspace is generated. This accumulation causes the Krylov subspace to diverge from the ideal one and prevents it from introducing new independent information. As discussed later in Sec.~\ref{sec:plain_lanczos}, the resulting convergence stalls at a much higher error level than expected. 

In this work, we focus on improving the convergence of MPS-based plain Lanczos methods (PLM) previously proposed in Refs.~\cite{dargel2012lanczos, huang2018generalized}, especially in the challenging case of targeting multiple eigenstates simultaneously. We first analyze why PLM fails to converge accurately for multiple target eigenstates, and then identify why the widely used thick-restart Lanczos method (TRL)~\cite{LehoucqSorensenYang1998, wu2000thick, wu1999thickrestart} does not resolve this issue when combined with the MPS representation: under MPS truncation, the residual directions associated with different Ritz vectors are no longer parallel. Consequently, a single restart vector is insufficient to provide the necessary correction information for all target Ritz pairs. To address this issue, we introduce a modified thick-block Lanczos (MTBL) scheme that restarts with a block formed by Ritz vectors and their associated residuals, which restores independent correction directions under truncation. This construction substantially improves the accuracy of Lanczos-type methods in the MPS representation while enabling to target several eigenstates simultaneously, thereby overcoming the main limitation that has hindered their broader use. Furthermore, by incorporating the shift-and-invert spectral transformation~\cite{wilkinson1965algebraic, ruhe1984rationalkrylov, golub2013matrix, saad2011largEeig, pietracaprina2018shiftinvert}, we can directly target excited states without first computing the lower-lying eigenstates, i.e., it does not suffer cumulative errors from the lower eigenstates. It can also be employed to refine approximate eigenstates obtained from other methods. 

We benchmark our algorithms on the half-filled Fermi–Hubbard chain and the spin-$1/2$ Heisenberg XXZ model in a longitudinal magnetic field. For both cases, MTBL significantly improves accuracy by three to seven orders of magnitude while remaining economical, compared to the previous Lanczos method within the MPS formalism. Notably, across these numerical experiments, the MTBL method reaches the accuracy limit imposed by the chosen bond dimension, achieving the best possible approximation of the target eigenstates allowed. This demonstrates that MTBL raises the accuracy of Lanczos-type methods to the level attainable by mainstream high-accuracy MPS-based approaches such as DMRG. To examine its scalability, we also apply our approach to a larger
Fermi-Hubbard model of size $L=50$ ($100$ spin sites) and Heisenberg XXZ chain of size $L=120$.

\section{Theoretical background}
\label{sec:background}

In this section, we review the classical Lanczos algorithm and the MPS/MPO formalisms, and then examine their integration. We also discuss the challenges posed by the MPS truncation, which motivates our subsequent improvement.

\subsection{The Lanczos algorithm} \label{lan}
The Lanczos algorithm is a widely used iterative method for computing the extremal eigenpairs of Hermitian matrices~\cite{haydock1980recursive, lanczos1950iteration, Paige1972, Cullum1985}, and has been widely used in diagonalizing large Hamiltonians within quantum many-body physics~\cite{laflamme2015efficient,weikert1996block}. Starting with an initial state $\ket{q_0}$, the Lanczos algorithm confines the evolving space to the Krylov subspace $\mathcal{K}_k$ spanned by $\{ \ket{q_0}, H\ket{q_0}, H^2 \ket{q_0},\dots, H^{k-1}\ket{q_0} \}$. The Lanczos procedure recursively constructs the Krylov vectors via a three-term recurrence relation:
\begin{subequations}
\label{eq:generate_krylov}
\begin{align}
\ket{w_j} &= H \ket{q_{j-1}}, \\
\ket{w_j'} &= \ket{w_j} - \ket{q_{j-1}} \braket{q_{j-1}|w_j} - \ket{q_{j-2}} \braket{q_{j-2}|w_j}, \label{eq:gram} \\
\ket{q_j} \, &= \ket{w_j'} / \norm{\ket{w_j'}}\label{eq:nor},
\end{align}
\end{subequations}
for  $j > 1$. The new Krylov vector $\ket{q_j}$ is formed by applying $H$ to $\ket{q_{j-1}}$, orthogonalizing it against the two previous Krylov vectors with the Gram-Schmidt process~\cite{PAECKEL2019167998} in Eq.~\eqref{eq:gram}, and finally normalizing the result by Eq.~\eqref{eq:nor}. For $j = 1$, one only needs to orthogonalize the vector against $\ket{q_0}$.

This iteration builds an orthonormal basis of the Krylov subspace $V = \text{span} \{\ket{q_0},\ket{q_1},\cdots,\ket{q_{k-1}}\}$. Projecting the original Hamiltonian $H$ onto the subspace $V$ yields a reduced (effective) Hamiltonian, whose elements are given by:
\begin{equation}
    \Tilde{H}_{ij} = \bra{q_i} H \ket{q_j}.
\end{equation} 
Diagonalizing this $k\times k$ matrix $\Tilde{H}$ yields the approximate eigenvalues $(\theta_i)_{i=0,\dots,k-1}$ together with the coefficient vectors $(\mathbf{y}_i)_{i=0,\dots,k-1}$. The approximation to the exact eigenvectors of the original Hamiltonian $H$ is reconstructed by back-transformation:
\begin{equation} 
\label{eq:RitzVectorDef} 
 \ket{\phi_i} =  \sum_{j=0}^{k-1} (\mathbf{y}_i)_j \ket{q_j}.
\end{equation}
The quantities $\theta_i$ and $\ket{\phi_i}$ are called the Ritz value and Ritz vector, respectively. After the set of Ritz pairs $\{ \theta_i,\ket{\phi_i} \}$ is obtained, each pair is tested for convergence; those that meet a desired tolerance of convergence are accepted as the numerical eigenpairs of $H$ \cite{bai2000templates, golub2013matrix}.

For many physically or chemically relevant Hamiltonians, only a few dozen or hundreds of basis vectors ($k \ll \dim(\mathcal{H})$) already yield accurate low‑lying eigenstates~\cite{haydock1980recursive, dagotto1994correlated, dargel2012lanczos}, and only these $k$ Krylov vectors need to be stored. The Lanczos algorithm turns a huge eigenvalue problem into a much smaller one inside the Krylov subspace, making it attractive for computational physics.

\subsection{Matrix product states and operators}
\label{sec:mps}

\begin{figure}
\centering
\begin{subfigure}{0.45\textwidth}
    \centering
     \begin{tikzpicture}[
    x=0.8cm, y=0.8cm,
    line width=1pt,
    every node/.style={inner sep=0pt,outer sep=0pt}
]
  \def\L{4}
  \def\shiftL{-0.2}
  \def\shiftEq{1.0}
  \def\shiftC{1.4}
  \def\edgeShift{0.30}
  \def\labelGapUp{0.15}
  \pgfmathsetmacro\ticksep{1 - 0.5*\edgeShift}
  \pgfmathsetmacro\yUpperTickEnd{0.8 + 0.4}

  \draw[rounded corners=1pt, fill=blue!40, draw=black]
        (\shiftL,0) rectangle (\L+\shiftL,0.8);
  \node at ({\shiftL + 0.5*\L},0.4) {$\ket{\psi}$};

  \foreach \i [evaluate=\i as \k using int(\i+1)] in {0,...,4} {
    \pgfmathsetmacro\xcoord{\shiftL + \edgeShift + \i*\ticksep}
    \ifnum\k=4
        \node[anchor=south,scale=1.2]
              at (\xcoord, \yUpperTickEnd + \labelGapUp) {$\dots$};
    \else
        \draw (\xcoord,0.8) -- (\xcoord,\yUpperTickEnd);
        \node[anchor=south,scale=1.2]
              at (\xcoord,\yUpperTickEnd + \labelGapUp)
              {\ifnum\k=5 $n_{L}$ \else $n_{\k}$ \fi};
    \fi
  }

  \node at ({\L+\shiftL+\shiftEq-0.5},0.4) {\large $=$};

  \foreach \i [evaluate=\i as \k using int(\i+1)] in {0,...,4} {
    \ifnum\k=4
        \node (A\i) at ({\L+\shiftL+\shiftC+\i},0.35) {\large$\cdots$};
    \else
        \node[
          draw, circle, fill=blue!40,
          minimum size=18pt, font=\scriptsize
        ] (A\i) at ({\L+\shiftL+\shiftC+\i},0.35)
          {\ifnum\k=5 $A[{L}]$ \else $A[\k]$ \fi};
        \coordinate (LegEnd\i) at ($(A\i.north)+(0,0.4)$);
        \draw (A\i.north) -- (LegEnd\i);
        \node[anchor=south,scale=1.2]
              at ($(LegEnd\i)+(0,\labelGapUp)$)
              {\ifnum\k=5 $n_{L}$ \else $n_{\k}$ \fi};
    \fi
  }

  \foreach \i/\j in {0/1,1/2,2/3,3/4} {
      \draw (A\i.east) -- (A\j.west);
  }
\end{tikzpicture}
    \caption{The wavefunction is obtained by contracting the ``tensor train" made of matrices $A[i]^{n_i}$.}
    \label{fig:mps}
\end{subfigure}
\hfill
\begin{subfigure}{0.45\textwidth}
    \centering
        \begin{tikzpicture}[
    x=0.8cm, y=0.8cm,
    line width=1pt,
    every node/.style={inner sep=0pt,outer sep=0pt}
]
\def\L{4}
\def\shiftL{-0.2}
\def\shiftEq{1.0}
\def\shiftC{1.4}
\def\edgeShift{0.30}

\def\yup{1.20}
\def\ydown{-0.40}
\def\labsep{0.15}
\def\labelGapUp{0.15}

\pgfmathsetmacro\crad{0.635/2}
\pgfmathsetmacro\ticksep{1 - 0.5*\edgeShift}
\pgfmathtruncatemacro\lastTick{\L+1}

\draw[rounded corners=1pt, fill=orange!40, draw=black]
      (\shiftL,0) rectangle (\L+\shiftL,0.8);
\node at ({\shiftL+0.5*\L},0.4) {$H$};

\foreach \i [evaluate=\i as \k using int(\i+1)] in {0,...,\L}{
  \pgfmathsetmacro\xcoord{\shiftL + \edgeShift + \i*\ticksep}

  \ifnum\k=\L
    \node[anchor=south,scale=1.2] at (\xcoord,\yup+\labelGapUp) {$\dots$};
  \else
    \draw (\xcoord,0.8) -- (\xcoord,\yup);
    \node[anchor=south,scale=1.2] at (\xcoord,\yup+\labelGapUp)
      {\ifnum\k=\lastTick $n_{L}$\else $n_{\k}$\fi};
  \fi

  \ifnum\k=\L
    \node[anchor=north,scale=1.2] at (\xcoord,\ydown-\labsep) {$\dots$};
  \else
    \draw (\xcoord,0) -- (\xcoord,\ydown);
    \node[anchor=north,scale=1.2] at (\xcoord,\ydown-\labsep)
      {\ifnum\k=\lastTick $m_{L}$\else $m_{\k}$\fi};
  \fi
}

\node at ({\L+\shiftL+\shiftEq-0.5},0.4) {\large $=$};

\foreach \i [evaluate=\i as \k using int(\i+1)] in {0,...,\L}{
  \pgfmathsetmacro\ycentre{\yup - 0.4 - \crad}

  \ifnum\k=\L
    \node (A\i) at ({\L+\shiftL+\shiftC+\i},\ycentre) {\large$\cdots$};
    \pgfmathsetmacro\xcoordRight{\L+\shiftL+\shiftC+\i}
    \node[anchor=south,scale=1.2] at (\xcoordRight,\yup+\labelGapUp) {$\dots$};
  \else
    \node[
      draw, circle, fill=orange!40,
      minimum size=18pt, font=\scriptsize
    ] (A\i) at ({\L+\shiftL+\shiftC+\i},\ycentre)
      {\ifnum\k=\lastTick $W[{L}]$\else $W[\k]$\fi};

    \draw (A\i.north)
          -- ($(A\i.north)!(0,\yup)!(A\i.north)$);
    \node[anchor=south,scale=1.2]
          at ($(A\i.north)!(0,\yup+\labelGapUp)!(A\i.north)$)
          {\ifnum\k=\lastTick $n_{L}$\else $n_{\k}$\fi};

    \draw (A\i.south)
          -- ($(A\i.south)!(0,\ydown)!(A\i.south)$);
    \node[anchor=north,scale=1.2]
          at ($(A\i.south)!(0,\ydown-\labsep)!(A\i.south)$)
          {\ifnum\k=\lastTick $m_{L}$\else $m_{\k}$\fi};
  \fi
}

\foreach \i/\j in {0/1,1/2,2/3,3/4}{
  \draw (A\i.east) -- (A\j.west);
}
\end{tikzpicture}
    \caption{MPO ansatz for an $L$-site chain, the operator can be reconstructed by contracting all the local matrices $W[i]^{m_i n_i}$.}
    \label{fig:mpo}
\end{subfigure}
\caption{Diagrammatic representation of matrix product states and operators.}
\end{figure}

Even though the Lanczos algorithm significantly reduces the complexity of the eigenvalue problem, it still runs into the curse of dimensionality issue since the memory needed to store the vectors (wavefunctions) explicitly scales exponentially with the system size.

Therefore, it is beneficial to employ the tensor network representation~\cite{ostlund1995thermodynamic, rommer1997class, schollwock2011density, Bridgeman_2017}, especially the matrix product state (MPS) ansatz, a widely used low-rank approximation of the wavefunction. Within the MPS ansatz, the wavefunction is rewritten as:
\begin{equation}
    \ket{\Psi} =
    \sum_{n_1, \dots, n_L} A[1]^{n_1} A[2]^{n_2} \cdots A[L]^{n_L} \ket{n_1, \dots, n_L }.
\end{equation}
Each inner tensor $A[i]$ ($1 < i < L$) is of order three, as illustrated in Fig.~\ref{fig:mps}. The superscript $n_i$ denotes the physical index labeling the possible states at site $i$, and for each $n_i$, the component $A[i]^{n_i}$ is a matrix of size $\chi_i \times \chi_{i+1}$. The parameter $\chi_i$ represents the bond dimension between sites $i - 1$ and $i$; in the following, we denote the maximum MPS bond dimension by $M$. The MPS representation offers a convenient ansatz for performing the low-rank approximation of the wavefunction. Controlled truncation of the MPS bonds via singular-value decomposition (SVD) can be applied directly to an existing MPS, a process justified by the Schmidt decomposition~\cite{schollwock2011density, hauschild2018efficient}, as depicted in Fig.~\ref{fig:compress_mps}. By employing such a truncation, the MPS ansatz can represent the quantum states more economically.

\begin{figure*}[t!]
\centering
\begin{tikzpicture}[
    x=1cm, y=1cm,
    font=\scriptsize,
    tensor/.style={circle, draw=black, fill=blue!30, minimum size=8mm, inner sep=1pt},
    tri/.style={draw=black, fill=cyan!40, regular polygon, regular polygon sides=3,
                regular polygon rotate=90, minimum size=10mm, inner sep=1pt},
    leg/.style={draw=black, line width=0.8pt},
    dottedleg/.style={leg, dash pattern=on 1pt off 2pt},
    midarrow/.style={
      decoration={markings, mark=at position 0.5 with {\arrow{>}}},
      postaction={decorate}
    }
]

\node[tensor] (A1) at (-3,0)    {$A[1]$};
\node[tensor] (A2) at (1.2-3,0)  {$A[2]$};
\node[tensor] (A3) at (2.4-3,0)  {$A[3]$};
\node[tensor] (AL) at (4.0-3,0)  {$A[L]$};
\draw[leg]       (A1.east) -- (A2.west);
\draw[leg]       (A2.east) -- (A3.west);
\draw[leg]       (A3.east) -- (3-3,0);
\draw[leg]       (3.45-3,0) -- (AL.west);
\draw[dottedleg] (3.1-3,0) -- (3.4-3,0);
\foreach \X/\lbl in {A1/1,A2/2,A3/3,AL/L} {
  \draw[leg] (\X.north) -- ++(0,0.3) node[above] [scale=1.5]{$m_{\lbl}$};
}

\Edge[Direct](2.2,0)(5,0)
\node at (4.45-1, 0.2)[scale=1.2]{Canonical form};

\node[tri] (R1) at (6,0)  [scale=0.7]  {$R[1]$};
\node[tri] (R2) at (7.8,0) [scale=0.7] {$R[2]$};
\node[tri] (R3) at (10.6-1,0) [scale=0.7] {$R[3]$};
\node[tri] (RL) at (12.4-1,0) [scale=0.7] {$R[L]$};
\draw[leg]       (R1.east) -- ++(0.6,0) -- (R2.west);
\draw[leg]       (R2.east) -- ++(0.6,0) -- (R3.west);
\draw[leg]       (R3.east) -- ++(0.1,0) -- (11.1-1,0);
\draw[leg]       (11.7-1,0) -- ++(0.1,0) -- (RL.west);
\draw[dottedleg] (11.25-1,0) -- ++(0.3,0) ;

\foreach \X/\lbl in {R1/1,R2/2,R3/3,RL/L} {
  \draw[leg] (\X.north) -- ++(0,0.4) node[above][scale=1.5] {$m_{\lbl}$};
}

\Edge[Direct](9.7,-0.6)(9.7,-1.9)
\node at (10.2,-1.2)[scale=0.9]{SVD};

\node[tri,rotate=180] (U1) at (6.5-0.3,-2.4) [scale=0.7] {\rotatebox{180}{$U[1]$}};
\node[rectangle, draw=black, fill=orange!60, rounded corners=1pt,
      minimum width=6mm, minimum height=6mm] (Smat)
    at (7.7-0.3,-2.4) {$\Sigma$};
\node[tri] (Vdag) at (8.9-0.3,-2.4) [scale=0.9] {$V^{\dagger}$};
\node[tri] (R20) at (10.1-0.3,-2.4) [scale=0.7]  {$R[2]$};
\node[tri] (R30) at (11.3-0.3,-2.4) [scale=0.7]  {$R[3]$};
\node[tri] (Rl0) at (12.5,-2.4) [scale=0.7]  {$R[L]$};

\draw[leg]       (U1.west) -- ++(0.32,0) -- (Smat.west);
\draw[leg]       (Smat.east) -- ++(0.35,0) -- (Vdag.west);
\draw[leg]       (Vdag.east) -- ++(0.35,0) -- (R20.west);
\draw[leg]       (R20.east) -- ++(0.35,0) -- (R30.west);
\draw[leg] (R30.east) -- ++(0.1,0);
\draw[leg]       (11.8,-2.4) -- (Rl0.west);
\draw[dottedleg] (11.45,-2.4) -- ++(0.3,0);

\Edge[Direct](5.8,-2.4)(3.2,-2.4)
\node at (4.45,-2.1)[scale=1.2]{Truncation};

\node at (-3.1,-2.1)[scale=1.2]{Continue SVD};
\node at (-3.2,-2.7)[scale=2]{...};
\Edge[Direct](-2,-2.4)(-4.3,-2.4)

\node[tri,rotate=180] (U1p) at (0.6-2,-2.4) [scale=0.7] {\rotatebox{180}{$U[1]$}};
\node[rectangle, draw=black, fill=orange!60, rounded corners=1pt,
      minimum width=6mm, minimum height=6mm] (Rp)
    at (1.8-2,-2.4) {$\Sigma$};
\node[tri] (R31) at (3.0-2,-2.4) [scale=0.7]  {$R[3]$};
\node[tri] (Rl1) at (4.2-1.4,-2.4) [scale=0.7]  {$R[L]$};

\draw[leg]       (U1p.west) -- ++(0.32,0) -- (Rp.west);
\draw[leg]       (Rp.east) -- ++(0.35,0) -- (R31.west);
\draw[leg]       (2.1,-2.4) -- (Rl1.west);
\draw[leg]       (R31.east) -- ++(0.1,0) ;
\draw[dottedleg] (1.6,-2.4) -- ++(0.3,0);

\end{tikzpicture}
\caption{The classical SVD method to compress the MPS. First, the MPS is brought into right-canonical form via successive QR decompositions. Then, sweeping from left to right, we carry out SVDs at each site and truncate the smallest singular values. Finally, the $\Sigma$ and $V^{\dagger}$ matrices are merged into the next site.}
\label{fig:compress_mps}
\end{figure*}

Alongside MPS, an analogous framework exists for operators: the matrix-product operator (MPO) formalism~\cite{schollwock2011density, keller2015efficient, chan2016matrix}:
\begin{equation}
\label{eq:MPO_def}
\begin{split}
    \hat{O} = \sum_{m,n} W[1]^{m_1 n_1} &W[2]^{m_2 n_2} \ldots W[L]^
    {m_L n_L} \\
    &\ket{m_1, \dots, m_L} \bra{n_1, \dots, n_L}.
\end{split}
\end{equation} 
For each site $i$, the local tensor $W[i]^{m_i n_i}$ carries two physical indices $( m_i, n_i)$ and can be viewed as a $\beta_i \times \beta_{i+1}$ matrix, where $\beta_i$ is the MPO bond dimension between sites $i - 1$ and $i$, as illustrated in Fig.~\ref{fig:mpo}. We denote the largest of these bond dimensions by $D$.  In Eq.~\eqref{eq:MPO_def}, the operator $\hat{O}$ is expressed as a summation of mappings from  $\ket{n_1,\dots,n_L}$ to $\ket{m_1,\dots,m_L}$, and the coefficients arise from contracting the chain of $W$ matrices. The MPO representation can fully utilize the sparsity of the physical or chemical Hamiltonian by constructing it properly~\cite{chan2016matrix, keller2015efficient, ren2020general, cakir2025optimal}. For example, the Heisenberg XXZ model can be represented exactly as an MPO of small bond dimension $D=5$~\cite{hauschild2018efficient}, regardless of the system size.

As with a Hamiltonian operator acting on a state vector to produce another vector, applying an MPO to an MPS yields a new MPS by contracting the physical indices at each site~\cite{schollwock2011density}. Likewise, adding two MPS produces a new MPS; both operations lead to boosted bond dimensions. To keep subsequent calculations feasible and efficient, especially in iterative algorithms, it is crucial to compress the resulting MPS, as illustrated in Fig.~\ref{fig:compress_mps}. There are also alternative MPS compression methods, including the zip-up method~\cite{Stoudenmire_2010}, the density matrix method~\cite{ma2024approximate}, and the variational method~\cite{schollwock2011density}, each offering different trade-offs in accuracy, efficiency, and applicability. In this work, we employ the standard SVD method, which is the most controllable and accurate.

\subsection{The Lanczos algorithm with MPS} 
\label{sec:plain_lanczos}

Since the Lanczos algorithm is inherently ``matrix-free'' and the MPS and MPO forms offer an economic way to represent Hamiltonians and states, it is reasonable to combine these tensor networks with the Lanczos algorithm~\cite{dargel2012lanczos}. 

As discussed before, the core step in the Lanczos algorithm is to compute new Krylov vectors via Eq.~\eqref{eq:nor}. In the tensor network representation, the term $H \ket{q_i}$ is realized by MPO-MPS multiplication, and the orthogonalization is implemented by MPS-MPS addition, leading to an enlarged MPS bond dimension. As discussed, the MPS compression is needed to reduce the bond dimension back to the capped one, introducing extra errors. 

The first consequence of this MPS truncation error is the loss of orthogonality. Because Krylov vectors are computed iteratively from preceding vectors, round-off errors accumulate and break global orthogonality. Repeatedly applying the Gram-Schmidt procedure does not solve this issue, as new errors arise during the MPS-MPS addition and subsequent compression steps. Ultimately, this is addressed by the re-orthogonalization technique proposed in~\cite{dargel2012lanczos}, which constructs a suitable linear combination of the current Krylov vectors so that the resulting vectors are well-orthogonalized:
\begin{equation}
    \ket{\psi_a} = \sum_{i = 0}^{a} C_{ai} \ket{q_i}
\label{eq:reortho}
\end{equation}
Note that we do not create new MPS for these orthogonal states $\{\ket{\psi_0}, \ket{\psi_1}, \ket{\psi_2}, \dots\}$; instead, we leave the original MPS Krylov vectors $\{\ket{q_0}, \ket{q_1}, \ket{q_2},\dots \}$ unchanged, only solve the matrix $C$ (by a Gram-Schmidt-like procedure or canonical orthogonalization method~\cite{jiang2021chebyshev}) and utilize $C$ in the following calculations. When calculating the effective Hamiltonian for the Lanczos algorithm, the matrix elements are calculated by:
\begin{equation}
    \Tilde{H}_{ab} = \bra{\psi_a} H \ket{\psi_b} = \sum_{i = 0}^{a} \sum_{j = 0}^{b} C^{*}_{ai} C_{bj} \bra{q_i} H \ket{q_j}.
\label{eq:eff_H}
\end{equation}
The Ritz pairs $\{ \theta_i,\ket{\phi_i} \}$ are obtained by solving the effective Hamiltonian $\Tilde{H}$, e.g., finding a real diagonal matrix $T$ and a unitary matrix $U$ so that $\Tilde{H} = U T U^{\dagger}$. The actual Ritz vector is obtained by:
\begin{equation}
    \ket{\phi_i} = \sum_{j=0}^{n-1} U_{ji} \ket{\psi_j} = \sum_{j=0}^{n-1} \sum_{k=0}^j U_{ji} C_{jk} \ket{q_k},
\label{eq:ritz}
\end{equation}
where $n$ is the size of the current Krylov subspace. Any desired quantity, especially the expectation value $\bra{\psi_i} \hat{O} \ket{\psi_i}$ for an operator $\hat{O}$, can be computed from the summation of the terms $\bra{q_i} \hat{O} \ket{q_j}$.

The second consequence of the MPS truncation error is the divergence from the desired Krylov space. Typically, only a small MPS bond dimension is needed to represent low-lying eigenstates, since these low-lying eigenstates obey an area law for entanglement. However, a larger bond dimension is required to represent $H\ket{\psi}$ if $\ket{\psi}$ is an arbitrary state, even when $\ket{\psi}$ itself can be approximated by an MPS with a small bond dimension~\cite{schollwock2011density, PAECKEL2019167998}. 
The reason is that the Hamiltonian $H$ is typically a sum of operators, and applying it to $\ket{\psi}$ can generate a more complex entanglement pattern, as each term introduces additional correlations that increase the overall entanglement in the resulting state. Therefore, the MPS truncation of $H\ket{\psi}$ generates a large error, and such an error prevents the Krylov vectors from ``exploring" the Hilbert space correctly; the resulting Krylov space deviates further from the intended one, and newly generated Krylov vectors no longer provide additional independent information in subsequent iterations. Consequently, as displayed later in Sec.~\ref{sec:benchmark}, the most significant result of this error is the severely broken convergence observed for most Hamiltonians. 

In this work, we focus on strategies to mitigate the impact arising from the second consequence of MPS truncation errors, i.e., improving the convergence of the Lanczos algorithms.

\section{MPS-based Lanczos Method with Block-Lanczos Concepts}

In this section, we first review the standard thick-restart approach~\cite{wu2000thick, wu1999thickrestart} and then illustrate that, when integrated with the block-Lanczos framework~\cite{golub1977block, saad1980rates, hackett2025blocklanczos}, the resulting modified Thick-Block Lanczos (MTBL) method markedly enhances the convergence of eigenstate calculations. For comparison, the original Lanczos procedure without any restart is referred to as the plain Lanczos method (PLM). We also introduce the shift-and-invert method, which reshapes the Hamiltonian spectrum, enabling the Lanczos algorithm to directly target the desired excited states. Additionally, we briefly discuss the use of energy variance as an indicator of accuracy for the approximate eigenstates.

\subsection{Thick-restart Lanczos Method}
\label{sec:thick}

Thick–restart Lanczos (TRL) method is one of the most widely used strategies to improve convergence and reduce memory usage in the Lanczos algorithm. In practice, it underpins mainstream software such as ARPACK~\cite{LehoucqSorensenYang1998} and numerous large–scale applications~\cite{wu2000thick, wu1999thickrestart}. In the TRL scheme, the iteration is interrupted after a $k$-step Lanczos process, and a thick set of $t$ Ritz vectors, which are approximate eigenpairs obtained from the Lanczos procedure, is retained together with the next Krylov vector $\ket{q_{k}}$. In the exact-arithmetic Lanczos method, the $\ket{q_{k}}$ vector is the residual vector for all the Ritz vectors, due to the three-term recurrence relation~\cite{Parlett1980, golub2013matrix}. Thus, such a single vector carries all the correction information for all the Ritz pairs. Concretely, the TRL exploits this property by restarting with:
\begin{equation}
[\,\ket{\phi_0},\dots,\ket{\phi_{t-1}},\, \ket{q_{k}}\,],
\end{equation}
and continue the Lanczos iteration from $\ket{q_{k}}$; we present the pseudo-code for classical TRL in Appendix~\ref{app:thick_table}. This construction preserves the solved spectral information, i.e., the kept Ritz pairs, while injecting exactly the direction that simultaneously reduces the residuals of all kept Ritz pairs. It can be shown that one cycle of TRL can simultaneously refine all $t$ targeted Ritz pairs~\cite{morgan1996restarting}.  

However, with MPS representations, the ideal Lanczos iterations are perturbed by the error induced by the MPS truncation. Consequently, the three-term recurrence relation does not hold anymore, and different Ritz pairs admit different residual vectors. In such a case, the next Krylov vector $\ket{q_{k}}$ cannot carry the correction information for all the Ritz vectors, and TRL thus fails, as presented in Sec.~\ref{sec:ferimi_benchmark}.

\subsection{Modified Thick-Block Lanczos Method}
\label{sec:mtbl}

Having identified the reason for the inaccuracy of the thick-restart scheme under the MPS representation, we are motivated to adopt the following strategy: since a single vector fails to capture the correction direction for all Ritz vectors, we compute the residual:
\begin{equation}
\label{eq:residual}
    \ket{r_i} = H\ket{\phi_i} - \theta_i \ket{\phi_i}
\end{equation}
corresponding to each Ritz vector. However, since the standard TRL can only employ a single correction direction, it cannot efficiently exploit all the residuals $\ket{r_i}$. To address this limitation, we adopt the block Lanczos method~\cite{hackett2025blocklanczos, golub1977block, saad1980rates}, which allows multiple vectors to serve as the initial vectors. 

\begin{algorithm}[H]
\caption{Modified Thick-Block Lanczos (MTBL) method}
\label{alg:trnl_mps}
\begin{algorithmic}[1]
\Statex \textbf{Input:}
\Statex \quad Hamiltonian $H$ as MPO, initial state $\ket{q_0} = \ket{\psi_0}$ as MPS
\Statex \textbf{Output:}
\Statex \quad $t$ eigenpairs $\mathcal{S}=\{(\theta_a,\ket{\phi_a})\}_{a=0}^{t-1}$ as MPS
\Statex \textbf{Note:}
\Statex \quad MPS is always compressed back to a given bond dimension $M$ and normalized after each operation

\State \textbf{Initialize} $\mathcal{S}=\emptyset$, $\mathcal{L}=\emptyset$, initial state $\ket{\psi_0}$

\State \textbf{Run} PLM with MPS representation with initial state $\ket{\psi_0}$. Retain the $t$ lowest Ritz pairs
$\{(\theta_a,\ket{\phi_a})\}_{a=0}^{t-1}$

\For{$i = 1,2,\dots,$ \textbf{until} convergence}

    \State \textbf{set} $Q_1 = [\, \text{Ritz vectors} \, \ket{\phi_{\ell}}\, \text{that are not in }\mathcal{L}\,]$

    \For{$\ket{\phi_a}\in Q_1$: $r_a = H\ket{\phi_a}-\theta_a \ket{\phi_a}$}, normalize $\ket{r_a}$
        \EndFor
        
    \State \textbf{Set} $Q_2=[\ket{r_{0}}, \ket{r_{1}}, \dots,\ket{r_{t-1}}]$
    
    \State \textbf{Set} $\mathcal{Q} = [Q_{1}, Q_{2}]$, orthogonalize vectors in $\mathcal{Q}$

    \For{$j = 2,3,\cdots$ \textbf{until} a pre-set number \textbf{or} convergence}
        \State $Q_{j+1} = H Q_j$
        \State Orthogonalize $Q_{j+1}$ to $Q_j$ and $Q_{j-1}$

        \State Append $Q_{j+1}$ to $\mathcal{Q}$
        \State Solve and update Ritz pairs $\{(\theta_a,\ket{\phi_a})\}$ with $\mathcal{L} \cup \mathcal{Q}$
    \EndFor

    \For{$\ell=0, 1, \dots, t - \abs{\mathcal{L} }- 1$} 
        \If{$(\theta_\ell,\ket{\phi_{\ell}}) \notin \mathcal{S}$ \textbf{and} $\norm{(H - \theta_\ell I)\ket{\phi_{\ell}}} < \epsilon$}
            \State \textbf{set} $\mathcal{S} = \mathcal{S} \cup \{(\theta_\ell, \ket{\phi_{\ell}})\}$, $\mathcal{L} = \mathcal{L} \cup \{\ket{\phi_{\ell}}\}$
        \EndIf
    \EndFor
    \If{$|\mathcal{S}| = t$} \State \textbf{break} \EndIf
\EndFor
\State \textbf{return} $\mathcal{S}$
\end{algorithmic}
\end{algorithm}

The original block Lanczos method begins with a block of $p$ orthonormal vectors $B_0=[\ket{q_{0}},\ket{q_{1}},\dots,\ket{q_{p-1}}]$ instead of a single vector $q_0$, and the Krylov basis is thus spanned by $\{ {B_0}, H{B_0}, H^2 {B_0},\dots, H^{k-1}{B_0} \}$. The block Lanczos method is also used in DMRG based on the bundled-MPS framework~\cite{baker2024direct}, whose efficiency has been discussed in Ref.~\cite{baker2024bundled} by T.E Baker and N. Seif. A related block-MPS strategy is employed in the multi-target update method (MTU)~\cite{lixuan2024mtu}, which likewise optimizes a block of MPSs simultaneously. The block Lanczos method is also applied in quantum computing~\cite{baker2024qc}.
In this work, we incorporate the block Lanczos with TRL when restarting the Lanczos procedure. Concretely, we retain two blocks of vectors, the first one contains the $t$ Ritz vectors:
\begin{equation}
    Q_1 = [\ket{\phi_0}, \ket{\phi_1}, \dots, \ket{\phi_{t-1}}],
\end{equation}
and the second one contains the $t$ residual vectors for these Ritz vectors:
\begin{equation}
    Q_2 = [\ket{r_0}, \ket{r_1}, \dots, \ket{r_{t-1}}],
\end{equation}
which are estimated by Eq.~\eqref{eq:residual}. The Lanczos procedure is restarted by repeating the following process:
\begin{enumerate}
    \item Generating a new block $Q_{i+1}$ by applying $H$ to  $Q_i$, and normalize the vectors in $Q_{i+1}$.
    \item Orthogonalize vectors in $Q_{i+1}$ to all the previous vectors by the Gram-Schmidt procedure.
\end{enumerate}
Finally, we solve the Ritz pairs using the re-orthogonalization discussed in Sec.~\ref{sec:plain_lanczos} and restart the Lanczos procedure until convergence or the desired accuracy is achieved. After each restart, any Ritz vector whose residual norm $\norm{(H - \theta_\ell I)\ket{\phi_\ell}} $ falls below the selected tolerance $\epsilon$ is locked, meaning that the corresponding Ritz vector $\ket{\phi_\ell}$ is regarded as converged and is excluded from subsequent restarts. Nevertheless, all newly generated Krylov vectors are orthogonalized against these locked vectors to ensure that no redundant components are regenerated, as detailed in Algorithm~\ref{alg:trnl_mps}.

In summary, even though both TRL and block Lanczos are classical Lanczos methods, our contribution is to adapt and combine these ideas in a way that specifically addresses the unparallelism induced by the MPS truncation. In particular, the classical TRL, which can improve the convergence of PLM, relies on the fact that the next Krylov vector represents the residual direction for all Ritz vectors, whereas this property no longer holds when MPS truncation exists. MTBL solves this issue by explicitly computing residual vectors for each individual Ritz vector, making the restart compatible with the MPS truncation.

\subsection{Shift-and-invert}
\label{sec:si}
In many applications, one is in fact more interested in excited states than in the ground state. It is therefore highly promising to directly focus on states around a given target energy, instead of successively converging to all lower-energy eigenstates. To achieve this, a method is to combine the Lanczos algorithm with a shift-and-invert spectral transformation~\cite{rano2025inexactlanczos}, where we replace the Hamiltonian $H$ with:
\begin{equation}
    B = (H - \lambda I)^{-1},
\end{equation}
where $\lambda$ is a shift chosen near the desired value of the spectrum, so that the eigenstates whose eigenvalues lie near $\lambda$ become extreme values of $B$. To construct the Krylov subspace of $B$, we need to evaluate
\begin{equation}
    \ket{w_j} = B\ket{q_{j-1}} = (H - \lambda I)^{-1} \ket{q_{j-1}}.
\end{equation}
However, it is generally ill-conditioned and also requires a prohibitively large bond dimension to represent $B$ as an MPO. To overcome this issue, we search for the MPS $\ket{w_j}$ which satisfies:
\begin{equation}
    \min\nolimits_{\ket{w_j}}
    \left\| (H - \lambda I)\ket{w_j} - \ket{q_{j-1}} \right\|_2^2.
\end{equation}
using the variational method~\cite{baiardi2019optimization, yu2017finding, rakhuba2016vibtt, pomata2023seeking}. By computing the extreme eigenstates of $B$, we directly obtain the desired excited states of $H$. For instance, the eigenstate of $H$ whose energy is closest to $\lambda$ from above corresponds to the largest eigenvalue of $B$. 

While the standard DMRG typically obtains excited states indirectly by solving all lower states sequentially, making its accuracy dependent on the accuracy of those preceding states, the Lanczos method with shift-and-invert can also refine the target excited state by using the DMRG result as the initial state. In this way, besides locating eigenstates near a chosen spectral position $\lambda$, the Lanczos method with shift-and-invert can further improve the precision of the targeted excited state without requiring convergence of all lower-energy states.

\subsection{Energy variance}
\label{sec:ev}

As the system size increases, reference eigenvalues from exact diagonalization become unavailable. In such cases, the accuracy of approximate eigenstates can be assessed using the energy variance, a widely applied diagnostic in computational physics and numerical linear algebra~\cite{golub2013matrix, cullum2002lanczos, martin2004electronic, hubig2018errorestimates, gleis2023controlled, gleis2022projector}. The energy variance of an approximate eigenpair $(\theta_i, \ket{\phi_i})$ with respect to a Hamiltonian $H$ is given by:
\begin{equation}
\sigma^2 = \bra{\phi_i} (H - \theta_i I)^2 \ket{\phi_i},
\label{eq:var}
\end{equation}
which measures how closely the approximate eigenvalue $\theta_i$ and eigenstate $\ket{\phi_i}$ approximate an exact eigenpair. Minimizing this variance is essential in iterative eigenvalue algorithms such as Lanczos and Davidson methods to achieve reliable convergence \cite{cullum2002lanczos}. 
To lower the computational cost of Eq.~\eqref{eq:var}, we should avoid explicitly forming $(H - \theta_i I)^2 \ket{\phi_i}$. Instead, we can evaluate the energy variance directly via tensor-network contractions, analogous to that employed in the DMRG algorithm~\cite{schollwock2011density, chan2016matrix}; in addition, a more efficient method for computing the energy variance in MPS form has also been developed~\cite{gleis2022projector}.

\section{Numerical Experiments}
\label{sec:benchmark}

To assess the performance of MTBL, we apply it to Hamiltonians of broad interest and demonstrate its accuracy. We also numerically confirm our physical intuition underlying the method's improvement. The total computational cost scales with the number of iterations, that is, with how many times the Hamiltonian $H$ is applied to the most recent Krylov vector (in PLM or TRL) or block (in MTBL) to generate the next one. That means, if the block size is $p$, one MTBL iteration involves $p$ MPO-MPS multiplication-compressions. In addition, MTBL needs extra $p$ such operations to compute the residual vectors at the beginning of each restart. For convenience, we use the iteration count as a direct measure of the computational cost, while we also provide the cost described in the number of MPO-MPS multiplication-compressions. The calculation with MPS representation is implemented in PyTeNet~\cite{pytenet}.

\begin{figure}
    \centering
    \includegraphics[width=0.9\linewidth]{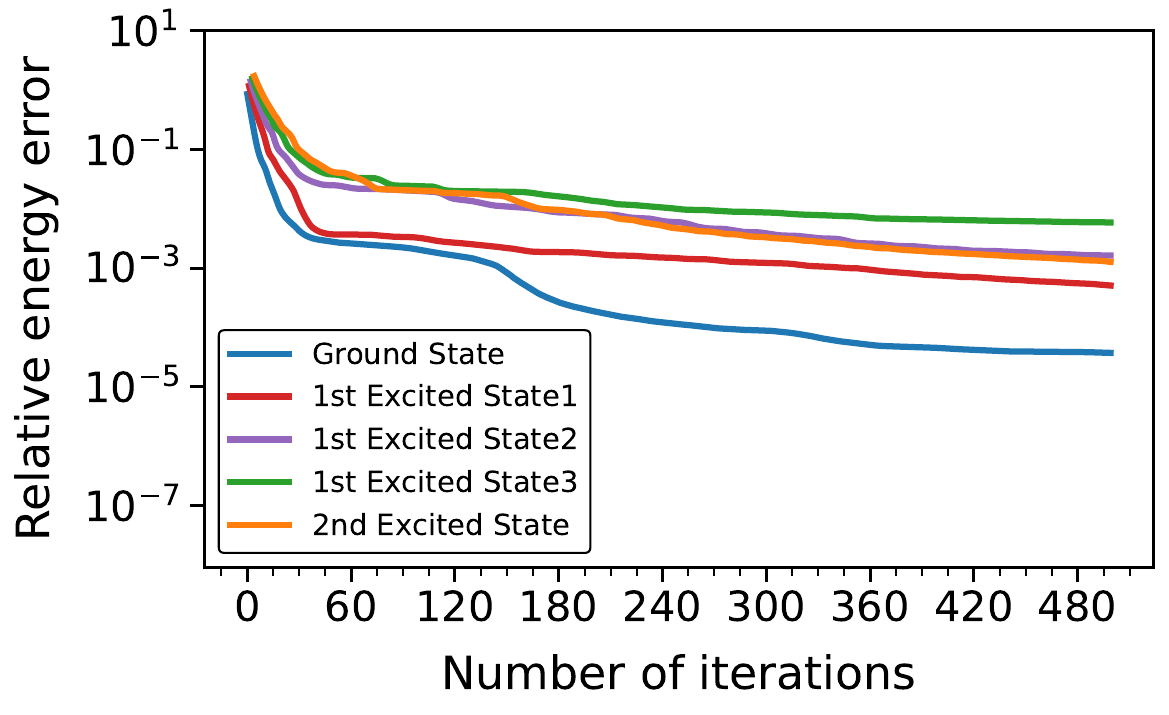}
    \caption{Convergence of the first five low-lying eigenstates of the Fermi–Hubbard model obtained with PLM in MPS representation. The excited states cannot converge to high accuracy. Here, each iteration refers to a single MPO-MPS multiplication-compression.}
    \label{fig:fermi_plain}
\end{figure}

\subsection{The Fermi-Hubbard model}
\label{sec:ferimi_benchmark}

We first benchmark the MTBL method using the one-dimensional Fermi–Hubbard model~\cite{Hubbard1963, Georges1996}, one of the most extensively studied systems in condensed matter physics. It describes the interaction of fermions on a lattice system, offering crucial insights into phenomena such as Mott insulating~\cite{Jordens2008, Hofrichter2016} and high-temperature superconductivity~\cite{Anderson1987, Scalapino1986, Dagotto1994}. The Fermi-Hubbard Hamiltonian is given by:
\begin{align}
H_{\text{FH}} = & -t \sum_{\langle i,j \rangle,\sigma}(a_{i\sigma}^{\dagger}a_{j\sigma} + a_{j\sigma}^{\dagger}a_{i\sigma}) \nonumber - \mu\sum_{i,\sigma} n_{i\sigma}\\
& + u\sum_{i}\left(n_{i\uparrow}-\frac{1}{2}\right)\left(n_{i\downarrow}-\frac{1}{2}\right)\nonumber,
\end{align}
where $a_{i\sigma}^{\dagger}$ and $a_{i\sigma}$ are fermionic creation and annihilation operators at site $i$ with spin $\sigma$, and $n_{i\sigma} = a_{i\sigma}^{\dagger}a_{i\sigma}$ denotes the number operator. We set the hopping amplitude to $t=1.0$, the chemical potential to $\mu=0$ (half-filling), and the on-site interaction to $u=8.0$. Such a parameter setting places the system in a strongly interacting regime, where correlation effects are expected to be significant. As a result, the accurate finding of low-lying eigenstates is more challenging than in weakly interacting regimes. The number of lattice sites is $L=8$ (i.e., 16 spin sites), so it is convenient to achieve the reference eigenpairs from exact diagonalization (ED)~\cite{dagotto1994correlated, sandvik2010computational}. We cap the Krylov vectors’ bond dimension at $M=32$, which means that we always compress the MPS bond dimension to $M=32$ after any operations, such as MPS-MPS addition or MPO-MPS multiplication. We initialize the calculation with a random state.

First, we present the energy errors for finding the first five low-lying eigenstates using the original plain Lanczos method (PLM). As illustrated in Fig.~\ref{fig:fermi_plain}, the convergence stalls far from the reference values, consistent with the discussion in Sec.~\ref{sec:plain_lanczos}, where we interpreted the PLM's failure to accurately find the eigenstates. 

\begin{figure}
    \centering
    \includegraphics[width=0.88\linewidth]{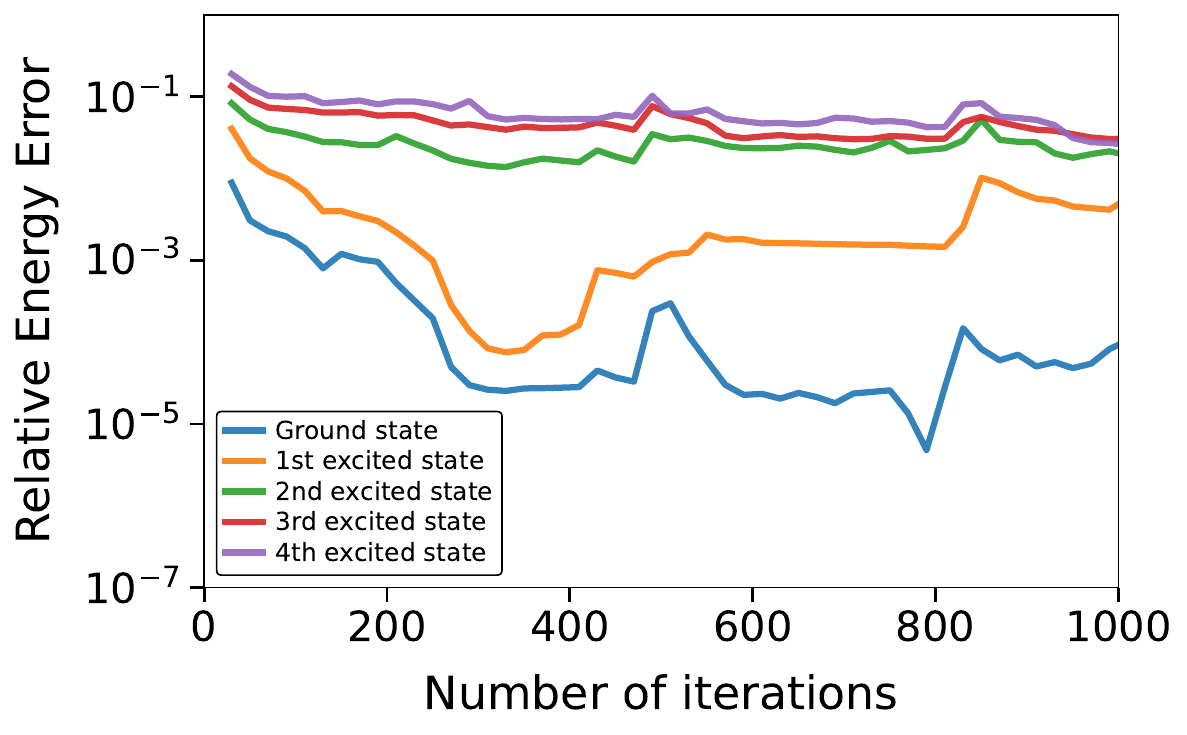}
    \caption{Convergence behavior of the first five low-lying eigenstates of the Fermi–Hubbard model using the TRL method with the MPS representation. We restart the procedure every 20 iterations, which corresponds to applying MPO-MPS multiplication-compressions for $20$ times. Only the ground state converges well since TRL using MPS can capture the residual vector of the ground state correctly.}
    \label{fig:fermi_trl}
\end{figure}

\begin{figure}
    \centering
    \includegraphics[width=0.85\linewidth]{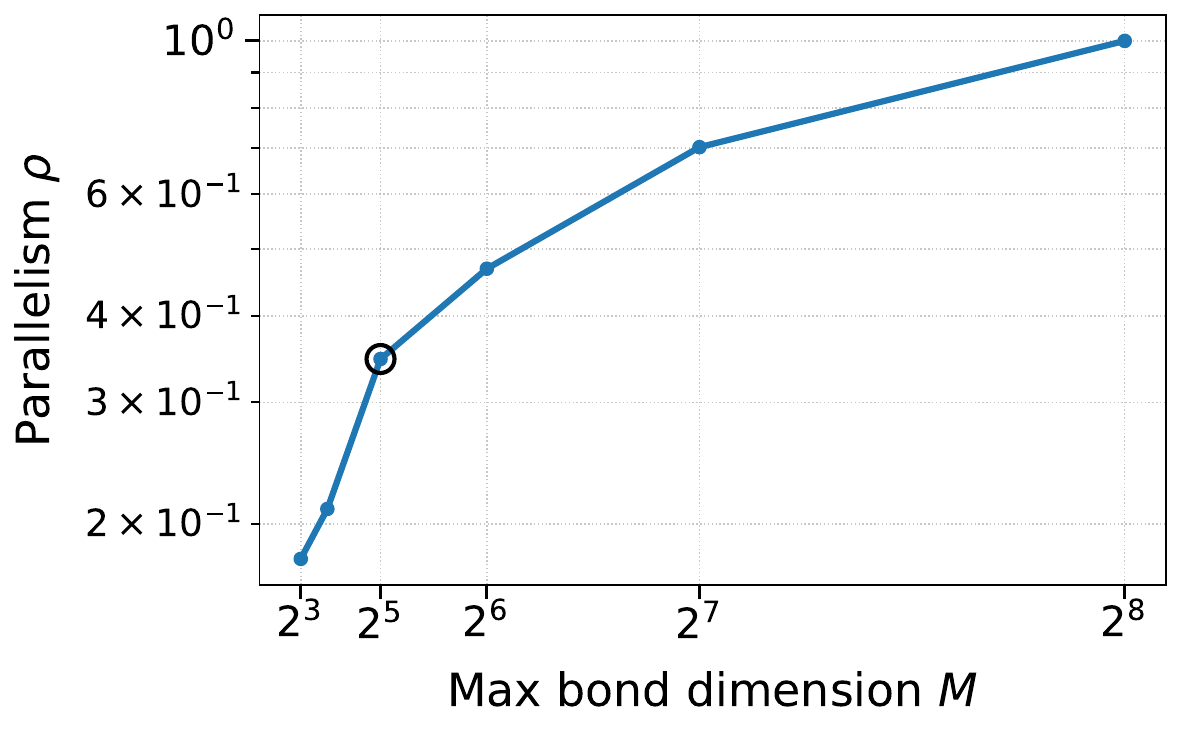}
\caption{Parallelism of the residual vectors associated with the five Ritz vectors for bond dimensions $[8,16,32,64,128,256]$, measured by $\rho$. The circled point corresponds to the case $M=32$, which leads to the overlap matrix Eq.~\eqref{eq:overlap_matrix}. Only when the bond dimension is set to $M=256$, which corresponds to the no-truncation case, are the residual vectors parallel.}
    \label{fig:trl_parallel}
\end{figure}

Classically, the TRL method is widely used to improve convergence. However, as discussed in Sec.~\ref{sec:thick}, it fails when the wavefunction is represented by an MPS ansatz, as illustrated in Fig.~\ref{fig:fermi_trl}. That is because the parallelism among the residual vectors of the corresponding Ritz vectors is lost due to MPS truncation, as discussed in Sec.~\ref{sec:thick}. Here, we provide numerical results that verify this conclusion. Specifically, we compute the residual vectors of the five lowest Ritz vectors obtained from PLM using Eq.~\eqref{eq:residual} and maximum bond dimension $M=32$, normalize each residual vector, and then evaluate the overlap matrix using these residuals as:
\begin{equation}
\label{eq:overlap_matrix}
    W_{ij} = \braket{r_i|r_j}= \begin{bmatrix}
1 & 0.3959 & 0.0074 & 0.0081 & 0.0152 \\
0.3959 & 1 & 0.1559 & 0.0174 & 0.0438 \\
0.0074 & 0.1559 & 1 & 0.2294 & 0.0965 \\
0.0081 & 0.0174 & 0.2294 & 1 & 0.1608 \\
0.0152 & 0.0438 & 0.0965 & 0.1608 & 1
\end{bmatrix}
\end{equation}
where we only show four decimal places. We see that these residual vectors are far from being mutually parallel, since perfect parallelism would yield an overlap matrix with all elements $1$. The parallelism of these residual vectors can be quantitatively measured by the following metric:
\begin{equation}
    \rho = \frac{\lVert W - I \rVert_{F}}{\sqrt{n(n-1)}} 
    = \sqrt{\frac{{\sum_{i \neq j} \lvert W_{ij}\rvert^2}}{{n(n-1)}}},
\end{equation}
where $\lVert \cdot \rVert_{F}$ is the Frobenius norm. $\rho=1$ indicates that the residual vectors are fully parallel, while $\rho=0$ corresponds to perfect mutual orthogonality. As shown in Fig.~\ref{fig:trl_parallel}, we present the parallelism of the residual vectors obtained from PLM under various bond dimensions, characterized by the metric $\rho$.

\begin{figure}
    \centering
    \includegraphics[width=0.9\linewidth]{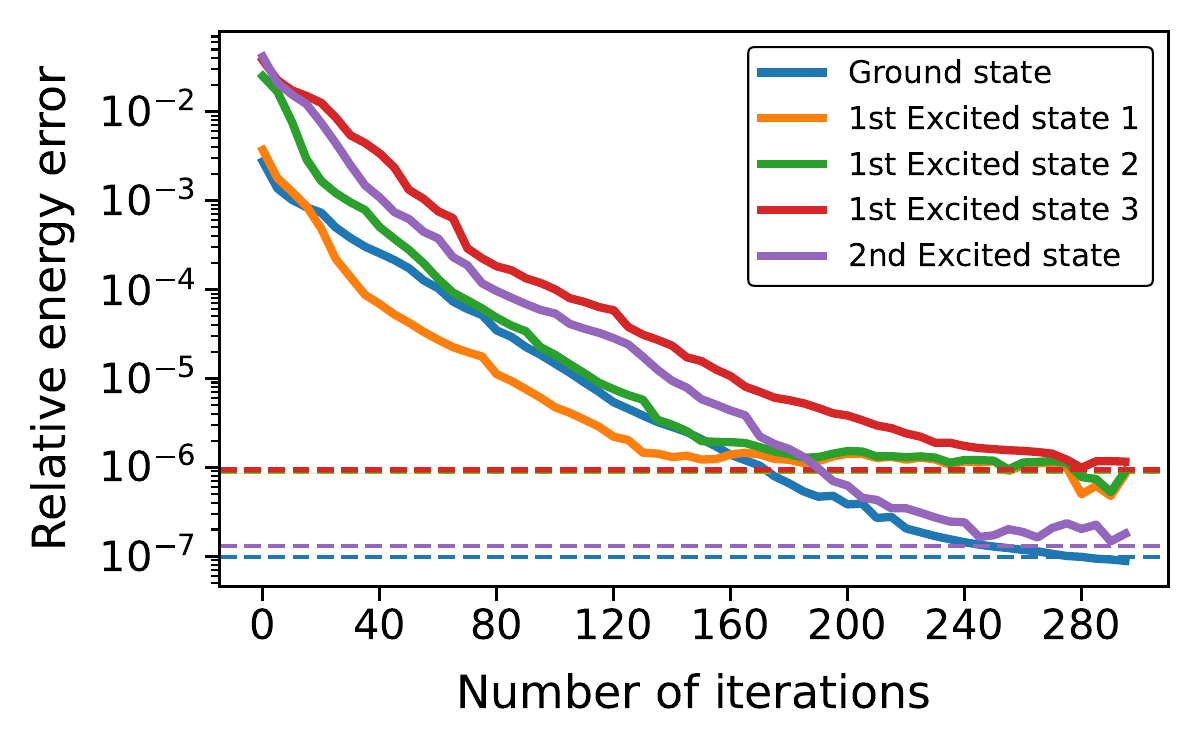}
    \caption{Track of the 5 lowest eigenstates of the Fermi–Hubbard model in the MTBL method. We use a block size of 6 and restart the procedure every 5 iterations, where each iteration contains 6 MPO-MPS multiplication-compressions, i.e., 36 each restart considering extra 6 such operations required to compute the residual vectors. For each eigenstate, the dashed line of the same color marks the optimal accuracy achievable with the chosen bond dimension. The degeneracy of the first three excited states is faithfully captured.}
    \label{fig:fermi_mrl}
\end{figure}

\begin{figure}
    \centering
    \includegraphics[width=0.85\linewidth]{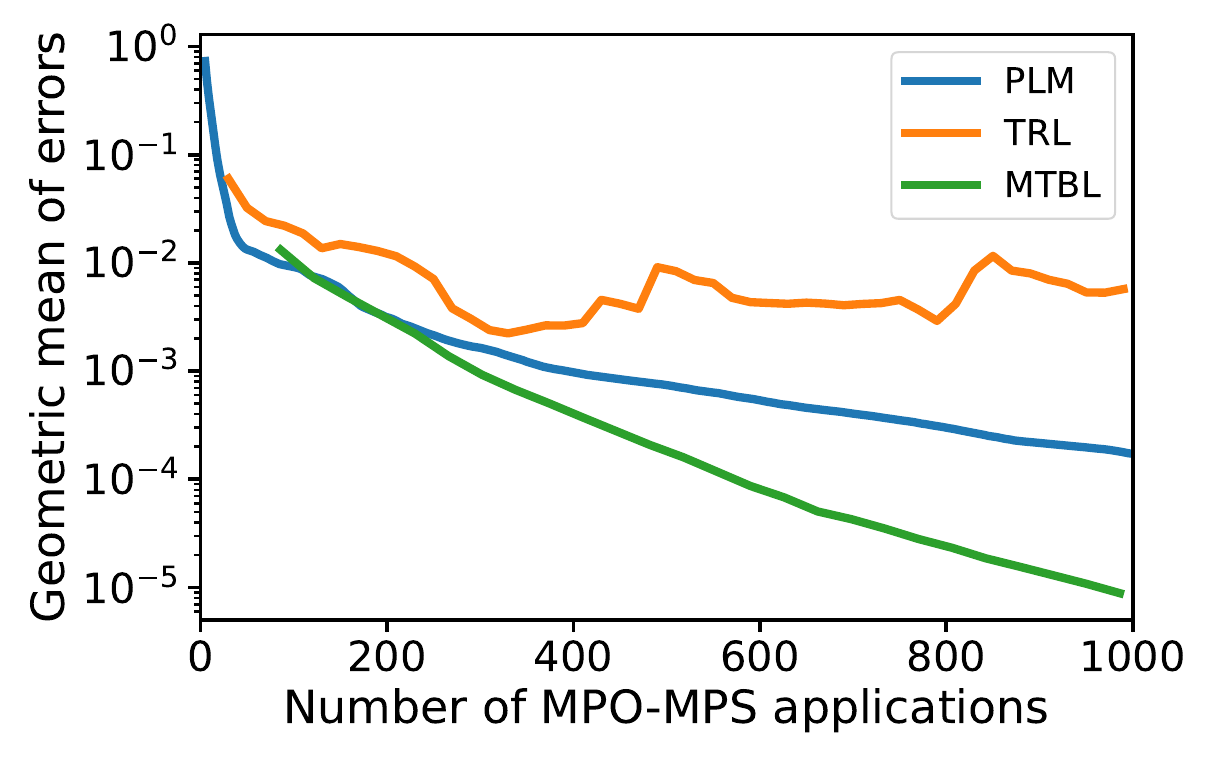}
    \caption{Convergence comparison of PLM, TRL, and MTBL using the geometric mean of relative energy errors, under the same computational cost metric by number of MPO-MPS applications.}
    \label{fig:compare_convergence_speed}
\end{figure}

\begin{table*}[t!]
    \centering
    \caption{Relative energy errors for the first five target eigenstates of the Fermi-Hubbard model, obtained by MTBL, DMRG with block Lanczos (BL), and MTU. MTBL yields the smallest relative energy error for each target eigenstate.}
    \label{tab:comparison_relative_errors_bundled}

    \vspace{0.6em}

    \setlength{\tabcolsep}{12pt}
    \renewcommand{\arraystretch}{1.25}
    \begin{tabular}{|l|c|c|c|}
        \hline
         & MTBL & DMRG w/ BL & MTU \\
        \hline
        Ground state        & $8.86250 \times 10^{-8}$ & $7.76557 \times 10^{-5}$ & $1.48308 \times 10^{-4}$ \\
        \hline
        1st excited state 1 & $8.45024 \times 10^{-7}$ & $1.91435 \times 10^{-5}$ & $9.19714 \times 10^{-5}$ \\
        \hline
        1st excited state 2 & $9.09718 \times 10^{-7}$ & $2.02649 \times 10^{-5}$ & $1.00167 \times 10^{-4}$ \\
        \hline
        1st excited state 3 & $1.15805 \times 10^{-6}$ & $5.97159 \times 10^{-5}$ & $2.15985 \times 10^{-4}$ \\
        \hline
        2nd excited state   & $1.79272 \times 10^{-7}$ & $3.02454 \times 10^{-5}$ & $2.35449 \times 10^{-4}$ \\
        \hline
    \end{tabular}
\end{table*}

Therefore, to fully exploit the residual information of all target eigenstates, we employ our MTBL, which brings all desired residual vectors into the restart procedure. Consequently, all target eigenstates converge in parallel and reach the optimal accuracy with the given bond dimension when employing MTBL to find the eigenstates, as shown in Fig.~\ref{fig:fermi_mrl}. Here, the optimal accuracy refers to the best approximation attainable with a given bond dimension. It is numerically estimated by solving the exact eigenstate in vector form through ED, converting it into an MPS, and truncating this MPS via the SVD method discussed in Sec.~\ref{sec:mps}; the accuracy of the resulting truncated MPS defines the optimal accuracy. All five target eigenstates reach a relative energy error of $10^{-5}$ within $160$ block iterations, corresponding to $960$ MPO-MPS multiplication-compressions. They further reach a relative energy error of $10^{-6}$ and attain the optimal accuracy within $280$ block iterations, corresponding to $2016$ such operations in total, or $403$ per target state. Compared to the PLM results shown in Fig.~\ref{fig:fermi_plain}, the accuracy is improved by three to four orders of magnitude.
Notably, the three degenerate first excited states are all accurately captured. In addition to the high accuracy, MTBL also converges faster compared to PLM and TRL. Using the number of MPO-MPS multiplication-compression as a unified metric of computational cost, we compare the convergence speeds of the plain Lanczos method (PLM), TRL, and MTBL for the Fermi-Hubbard model using the geometric mean of the relative energy errors. Specifically, for each method, we define the averaged relative error over the target eigenstates as
\begin{equation}
    \epsilon_g =
    \left( \prod_{i=1}^{n_{\mathrm{tar}}} \epsilon_i \right)^{1/n_{\mathrm{tar}}},
\end{equation}
where $\epsilon_i$ denotes the relative energy error of the $i$th target eigenstate, and $n_{\mathrm{tar}}$ is the number of target eigenstates. As shown in Fig.~\ref{fig:compare_convergence_speed}, the averaged error of MTBL decreases to nearly $10^{-4}$ with 400 MPO-MPS multiplication-compressions, whereas those of TRL and PLM remain around $10^{-3}$. This demonstrates that MTBL not only achieves higher accuracy but also converges faster under the same cost metric. Furthermore, we also observe that during the MTBL simulation, the MPO-MPS compression error does not necessarily increase with a fixed bond dimension. We discussed this in Appendix~\ref{app:compression_error}.

Here, we further compare MTBL with methods based on the bundled MPS. We implemented two representative methods: the DMRG algorithm with block Lanczos (BL)~\cite{baker2024direct}, and the multi-target update method (MTU)~\cite{lixuan2024mtu}. We used the same parameter settings for the Fermi-Hubbard Hamiltonian as above and capped the bond dimension at $M=32$. We computed the lowest five eigenstates and evaluated the relative energy errors. As summarized in Table~\ref{tab:comparison_relative_errors_bundled}, MTBL achieves substantially smaller relative energy errors than the two bundled-MPS-based methods for all target eigenstates.

\begin{figure}
    \centering
    \includegraphics[width=0.9\linewidth]{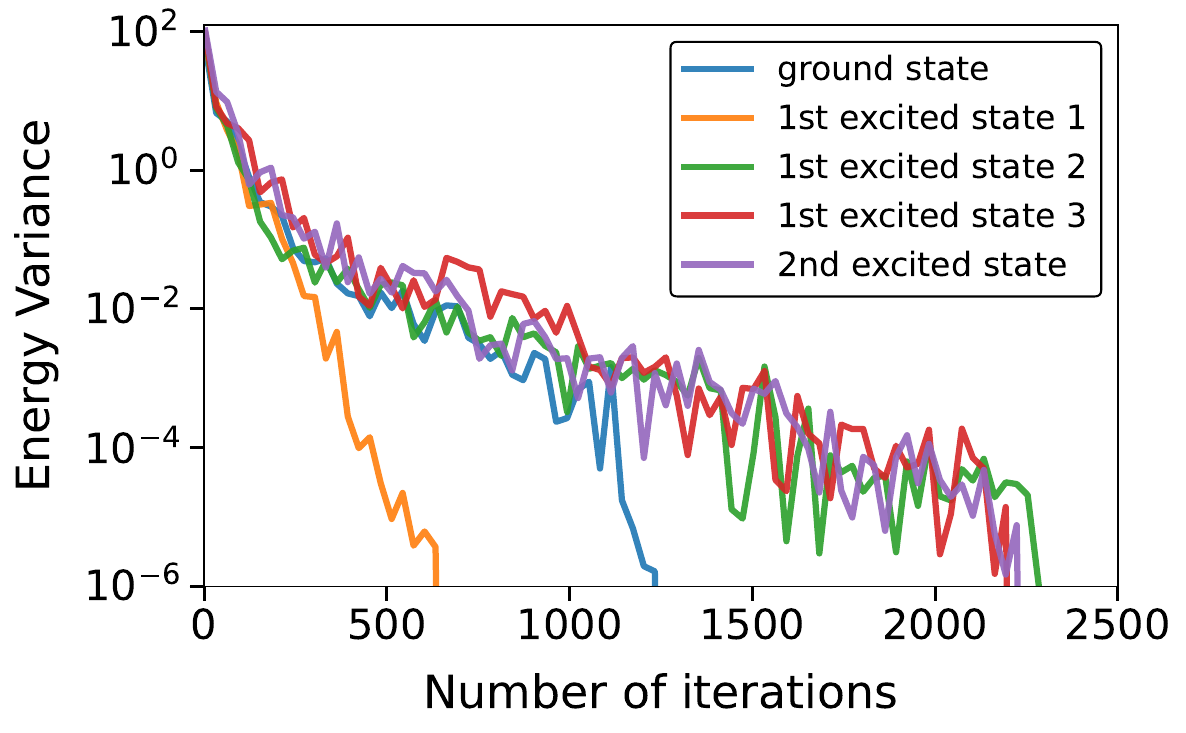}
    \caption{Track of the 5 lowest eigenstates of the Fermi–Hubbard model with $L=50$ in the MTBL method. We use a block size of 6 and restart the procedure every 3 iterations, where each iteration contains 6 MPO-MPS multiplication-compressions, i.e., 24 each restart. The degeneracy of the first three excited states is confirmed by checking the Ritz values.}
    \label{fig:fermi_hubbard_L50}
\end{figure}

To demonstrate the scalability and reliability of MTBL, we extend the size of the Fermi-Hubbard model to $L=50$, which corresponds to $100$ spin sites and Hilbert space of dimension $2^{100}$; we cap the maximum bond dimension at $M=300$. We lock a target state once its energy variance reaches $10^{-6}$. After being locked, this state is excluded from subsequent Lanczos iterations and is no longer updated. However, the locked states are still included when solving the effective Hamiltonian, so that the higher eigenstates remain orthogonal to them. As shown in Fig.~\ref{fig:fermi_hubbard_L50}, all five target states reach the target energy variance of $10^{-6}$, each requiring approximately 2900 MPO-MPS multiplication-compressions on average.

\subsection{The spin-$1/2$ Heisenberg XXZ model in a longitudinal magnetic field}
The second model we chose is the one-dimensional spin-$1/2$ Heisenberg XXZ model~\cite{Heisenberg1928, YangYang1966I, YangYang1966II, Alcaraz1987, Bertini2016} in a uniform longitudinal magnetic field: 
\begin{align}
H_{\text{XXZ}} = &\, J \sum_{\langle i,j \rangle} \frac{1}{2}(S_i^{+} S_j^{-} + S_i^{-} S_j^{+}) 
+ \Delta \sum_{\langle i,j \rangle} S_i^{z} S_j^{z} \nonumber \\
& - h \sum_i S_i^{z}.
\end{align}
We denote $S_i^\pm = S_i^x \pm i\,S_i^y$ operators as raising/lowering operators, and $S_i^\alpha = \tfrac{1}{2}\sigma_i^\alpha\ \ (\alpha=x,y,z)$ where $\sigma_i^\alpha$ are the Pauli $\alpha$ matrices. We set the nearest-neighbor exchange coupling strength to $J=-4$, the anisotropy parameter to  $\Delta = -4$, and the uniform magnetic field strength to $h=2$. The number of lattice sites is $L=16$; we cap the Krylov vectors’ bond dimension at $M=2$. We work in the total spin $S^z_{\text{tot}} = 1$ sector since $h > 0$ favors positive magnetization, and the initial state is set as a random state satisfying this total spin.

\begin{figure}[b!]
    \centering
    \includegraphics[width=0.9\linewidth]{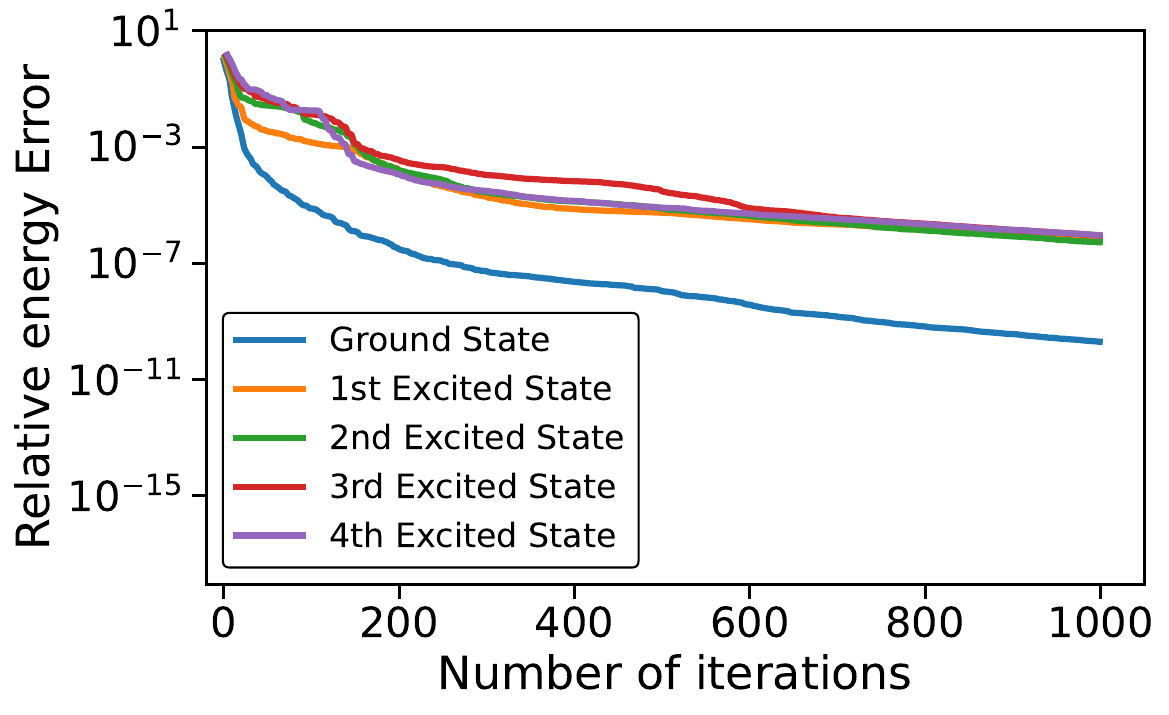}
    \caption{When using the PLM to simulate the Heisenberg model, the convergence stalls far away from the reference eigenvalues, except for the ground state. Here, each iteration refers to a single MPO-MPS multiplication-compression.}
    \label{fig:hei_plain}
\end{figure}

\begin{figure}[b!]
    \centering
    \includegraphics[width=0.9\linewidth]{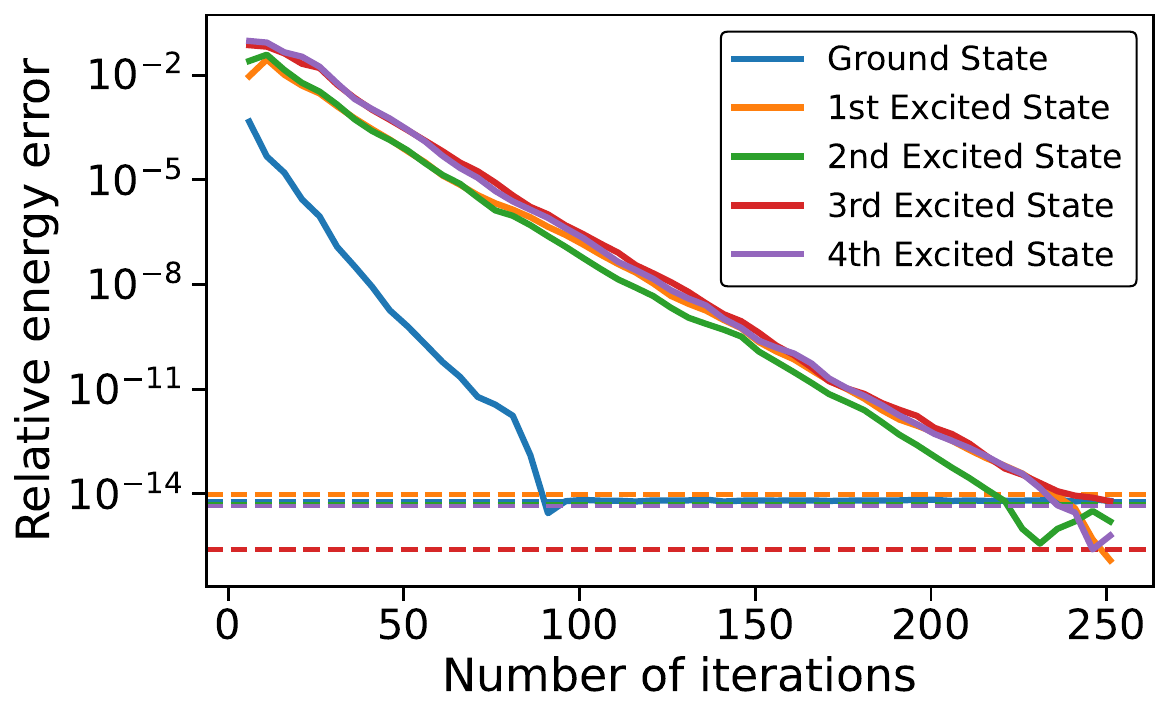}
    \caption{Convergence of the five relative energy errors of the low-lying eigenstates of the Heisenberg model ($L=16$) obtained using the MTBL method. Each block contains 5 vectors, and we restart the procedure every 5 iterations, for simplicity. Here, each iteration contains 5 MPO-MPS multiplication-compressions, and this indicates that we restart after every 30 such operations.
    For each eigenstate, the dashed line of the same color indicates the optimal accuracy achievable with the chosen bond dimension. The bond dimension is capped at $2$.}
    \label{fig:hei_mrl}
\end{figure}

\subsubsection{Finding low-lying states using MTBL}

Similar to the previous example, the PLM again fails to accurately recover the low-lying eigenstates. As shown in Fig.~\ref{fig:hei_plain}, even though the eigenstates appear to converge, the subspace size has grown to $1000$, corresponding to 1000 MPO-MPS multiplication-compressions, which is already prohibitively large. Such an oversized subspace not only makes the calculation and diagonalization of the effective Hamiltonian expensive but also leads to a cascade of issues, including numerical instability, increased memory consumption, and loss of orthogonality. In contrast, the MTBL method achieves rapid and stable convergence, as displayed in Fig.~\ref{fig:hei_mrl}. The relative energy errors drop below $10^{-8}$ within $150$ block iterations, and all five eigenstates are successfully obtained, reaching their optimal accuracies in fewer than $250$ block iterations. Equivalently, reaching a relative energy error below $10^{-8}$ requires $900$ MPO-MPS multiplication-compressions in total, corresponding to $180$ such operations per target state.
To demonstrate the robustness of MTBL, we also benchmark MTBL with the parameter settings $J=1$, $\Delta=1$, and $h=1$. 
For this parameter setting, we test MTBL with two system sizes: $L=16$ with $M=16$, as well as $L=100$ with $M=300$.
To keep the main text concise, we present the results in Appendix~\ref{app:benchmark_heisenberg}, where MTBL also achieves the optimal accuracy.

\subsubsection{Directly targeting excited states}
\begin{figure}
    \centering
    \includegraphics[width=0.9\linewidth]{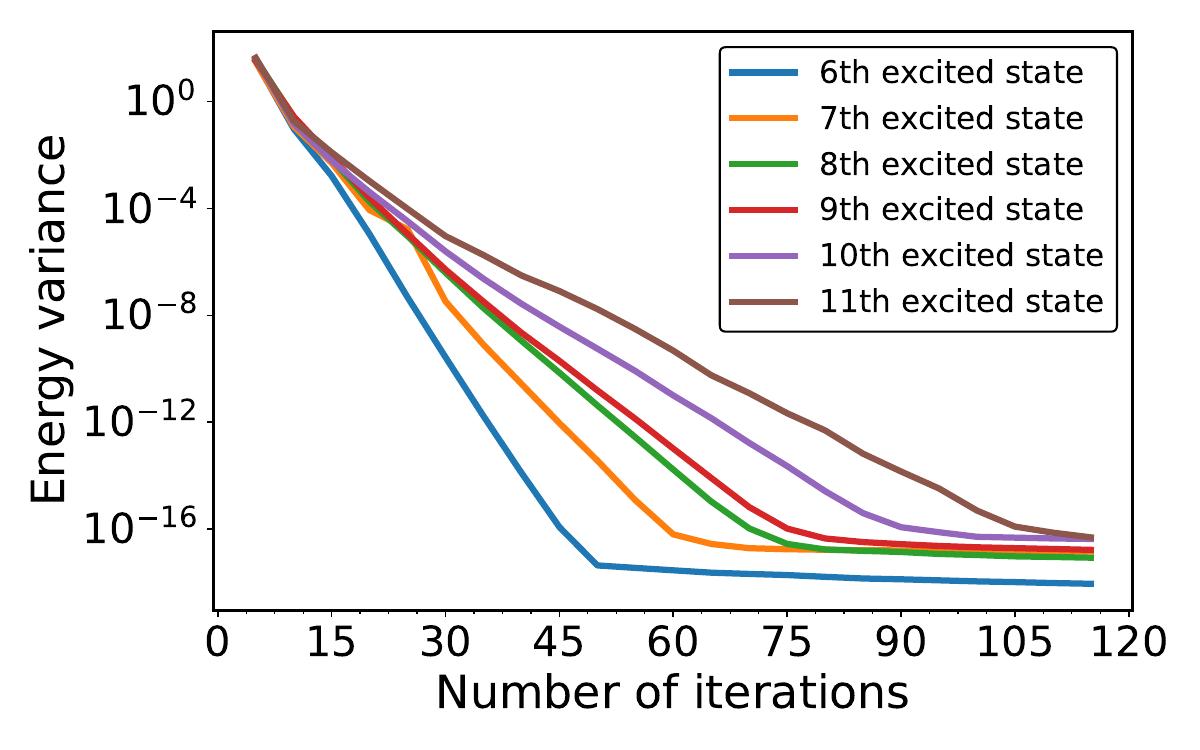}
    \caption{Direct targeting of excited states close to $\lambda$ in the spectrum using the shift-and-invert technique. Each block contains 6 vectors, and we restart the procedure every 6 iterations, i.e., $42$ MPO-MPS multiplication-compressions. We assess accuracy using the energy variance, as we do not know the corresponding eigenstates before comparing them with the reference after convergence. The bond dimension is capped at $4$.}
    \label{fig:si}
\end{figure}

We further demonstrate that, by employing the shift-and-invert technique, we can directly target excited states. Using the same parameter settings as above, we choose an energy shift of $\lambda=-27.5$ and compute six eigenstates whose energies are larger than this shift, which corresponds to the six highest positive eigenstates of $B$. After the convergence of these target eigenstates, we compare them with the reference spectrum and identify them as the 6th–11th excited states. As shown in Fig.~\ref{fig:si}, all these eigenstates converge to an energy variance below $10^{-15}$ within 120 iterations, or 840 MPO-MPS multiplication-compressions, equivalently. Also, eigenvalues closer to the shift $\lambda$ converge more rapidly, and this behavior is consistent with the spectral transformation induced by shift-and-invert: for an eigenpair $(\theta_i,\ket{\phi_i})$ of $H$, the corresponding eigenvalue of $B = (H - \lambda I)^{-1}$ is $\mu_i = \frac{1}{\theta_i-\lambda}.$
Hence, the closer $\theta_i$ is to the shift $\lambda$, the larger $|\mu|$ becomes, making the associated state more extreme in the spectrum of $B$ and therefore typically converging faster in the Lanczos iteration.

\subsubsection{Benchmark on large system $L=120$}

\begin{figure}
    \centering
    \includegraphics[width=0.9\linewidth]{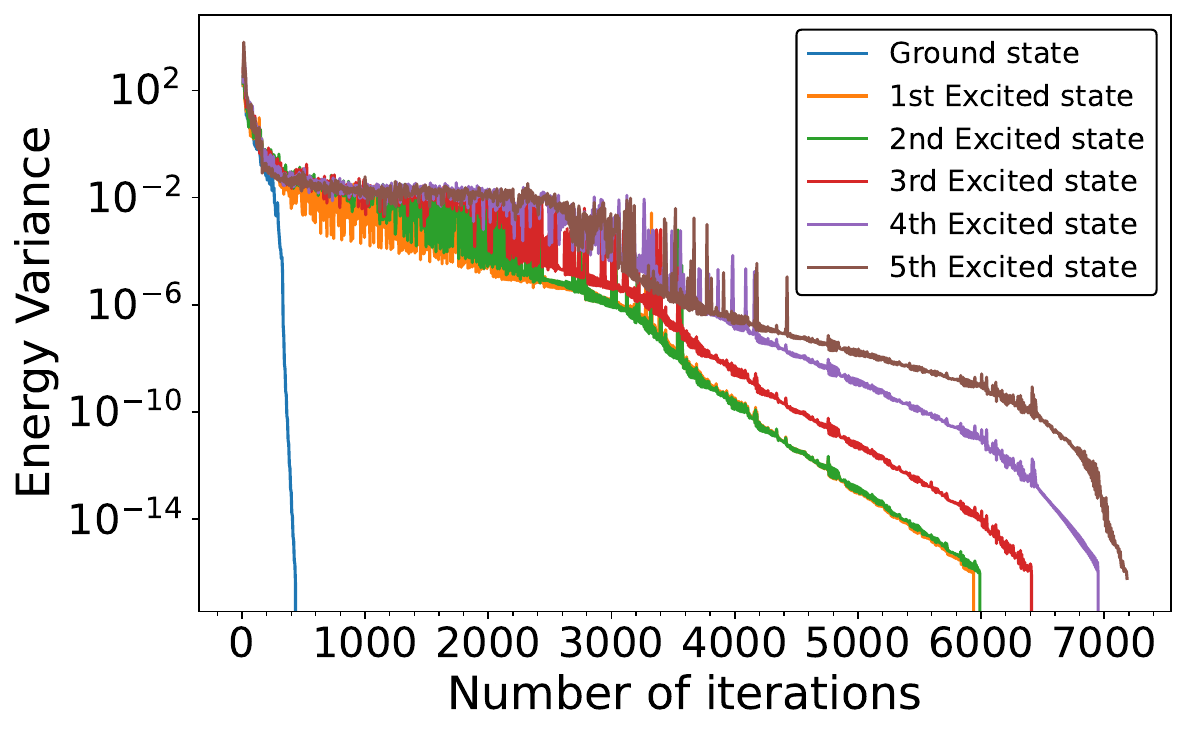}
    \caption{Convergence of the MTBL method for computing the six lowest eigenstates of the 120-site Heisenberg XXZ model. Each block contains 6 vectors, and we restart the procedure every 5 iterations, i.e., $36$ MPO-MPS multiplication-compressions. More iterations are required because a random initial state has only a small overlap with the low-lying eigenstates. The number of iterations can be reduced by selecting a larger bond dimension.} 
    \label{fig:xxz_120}
\end{figure}
We also demonstrate the scalability of MTBL with a larger Heisenberg chain. We increase the system size to $L=120$, which cannot be solved by exact diagonalization because of memory limitations, while restricting the bond dimension of the Krylov vectors to $M=10$. As discussed in Sec.~\ref{sec:ev}, we will use the energy variance as the metric of errors since the exact reference values are not available from ED for such a large system. As shown in Fig.~\ref{fig:xxz_120}, all six target states converge to energy variances below $10^{-14}$. These results demonstrate that the MTBL method is both scalable and reliable for computing low-lying eigenstates of large many-body systems. As the energy variance reflects how close the Ritz vectors are to the exact eigenstates, this indicates that all target states have converged to a very high accuracy.

\section{Conclusion and Discussion}

In this work, we propose targeted improvements to the Lanczos method within the MPS representation. The MTBL method leverages the block-Lanczos framework to capture the residual directions of all Ritz vectors, which are no longer parallel under the MPS ansatz. The block-Lanczos scheme serves as a key component that enables the efficient propagation of more information across successive iterations. This residual-augmented restart distinguishes MTBL from previous MPS-based Lanczos methods, whereas earlier approaches continue with either a single Ritz vector or a single residual direction. We demonstrated the accuracy, reliability, and scalability of our method by benchmarking on the Fermi–Hubbard model and the Heisenberg model in a longitudinal field. We expect the MTBL to provide a reliable eigenstate solver that inherently avoids local minima, decouples the accuracy of excited states from lower-lying eigenstates when combined with the shift-and-invert technique, and naturally resolves degeneracies through a block formulation. As discussed in Ref.~\cite{rano2025inexactlanczos, larsson2026accurate}, the Lanczos method combined with shift-and-invert provides a powerful alternative for finding eigenstates within an energy window that has a high density of eigenstates.

To demonstrate the convergence of our MTBL method, we employ the standard singular value decomposition (SVD) for MPS compression. Nevertheless, alternative compression strategies can also be considered, such as the zip-up method~\cite{Stoudenmire_2010}, the density-matrix-based approach~\cite{ma2024approximate}, the variational optimization scheme~\cite{schollwock2011density}, and the successive randomized compression (SRC) method~\cite{camano2025src}. Each compression technique offers a distinct balance between computational cost, memory usage, and accuracy. It is therefore promising to explore these alternative schemes to further accelerate MTBL, potentially at the expense of some accuracy.

Because the MTBL only operates on MPO-MPS multiplication, MPS-MPS addition, and compression, it can be integrated into any modern tensor network library without requiring additional heavy infrastructure. Additionally, it is not confined to the MPS/MPO framework; any ansatz supporting these operations, e.g., projected entangled pair states (PEPS)~\cite{Verstraete2006PEPS}, isometric tensor-network states (isoTNS)~\cite{Zaletel2020isoTNS}, tree tensor network states (TTNS)~\cite{Shi2006TTNS}, and so on, can be incorporated into our method.

\begin{acknowledgements}
We thank Chao Yang, Alec Dektor, Roel Van Beeumen, and Jan von Delft for helpful discussions. We also acknowledge funding by the Munich Quantum Valley initiative, supported by the Bavarian state government with funds from the Hightech Agenda Bayern Plus. 
\end{acknowledgements}

\section*{Author Contributions}

Y.W. conceived the project, proposed the main idea, developed the initial MPS-Lanczos code, performed the numerical experiments, and wrote the manuscript. Z.Y. contributed to the conception of the project, developed the MTBL code, and performed numerical experiments. X.W. implemented the shift-and-invert method and performed the benchmark comparisons among MTBL, MTU, and DMRG with the block Lanczos algorithm. C.B.M. supervised the project, participated in the discussion, and revised the manuscript.

\section*{Data Availability}
The data that support this study are available on Zenodo~\cite{datas}. The codes are available from the corresponding authors upon reasonable request.

\appendix
\section{Process of Thick-restart Lanczos Method} \label{app:thick_table}

In this appendix, we summarize the practical steps of the Thick-Restart Lanczos (TRL) algorithm in the exact-arithmetic Lanczos method.

The TRL improves convergence and reduces memory usage by periodically restarting the Lanczos iteration with preserved spectral information from the previous iteration. This implementation serves as the foundation for our MTBL in the main text. Additionally, the Ritz vector that reaches the desired accuracy can also be locked in TRL to save computational resources.

\begin{algorithm}[H]
\caption{Thick-Restart Lanczos (TRL)} \label{alg:trl}
\begin{algorithmic}[1]
\Statex \textbf{Input:}
\Statex \quad Hermitian matrix $H$, initial state $\ket{q_0} = \ket{\psi_0}$, subspace size $k$, error tolerance $\epsilon$
\Statex \textbf{Output:}
\Statex \quad $t$ lowest eigenpairs $\mathcal{S}=\{(\theta_i,\ket{\phi_i})\}_{i=0}^{t-1}$

\State Run the first cycle Lanczos algorithm with initial state $\ket{q_0}$ and Krylov subspace size $k$ to obtain the Krylov subspace $Q=\{\ket{q_0}, \ket{q_1}\dots,\ket{q_{k-1}}\}$.
\State Compute the next Krylov vector $\ket{q_k}$
\State Compute Ritz pairs 
$(\theta_i,\ket{\phi_i})$ with Krylov subspace $Q$

\For{$i = 1,2,\dots,$ \textbf{until} convergence}

    \State Set $\ket{q_n} = \ket{\phi_n}$ for $n = 0, 1,\dots, t-1$
    \State Set $\ket{q_t} = \ket{q_k}$
    \State Orthogonalize $\ket{q_t}$ against previous vectors
    \State Initialize Lanczos procedure with $\{\ket{q_0}, \ket{q_1}\dots,\ket{q_{t}} \}$
    \State Extend new Krylov vectors until subspace size $k$
    \State Define $Q=\{\ket{q_0}, \ket{q_1}\dots,\ket{q_{k-1}}\}$
    \State Compute an extra Krylov vector $\ket{q_{k}}$
    \State Update Ritz pairs $(\theta_i,\ket{\phi_i})$ with $Q$
    \State Sort Ritz pairs $(\theta_i,\ket{\phi_i})$ in ascending order of $\theta_i$
    \For{$\ell=0,1,\dots,t-1$} 
    \State Compute $\epsilon_\ell = \|(H-\theta_\ell I)\ket{\phi_\ell}\|$ 
    \EndFor
    \If{$\epsilon_\ell < \epsilon$ for all $\ell=0,1,\dots,t-1$}
    \State Stop: convergence achieved
    \State \textbf{set} $\mathcal{S} = \{(\theta_\ell, \ket{\phi_{\ell}})\}$ for all $\ell=0,1,\dots,t-1$
    \EndIf
\EndFor
\State \textbf{return} $\mathcal{S}$
\end{algorithmic}
\end{algorithm}

\section{The MPO-MPS truncation error during MTBL simulation}
\label{app:compression_error}

Controlling the bond dimension is a central issue in MPS-based simulations. In real-time evolution, the entanglement entropy of
\begin{equation}
    \ket{\psi(t)} = e^{-itH}\ket{\psi}
\end{equation}
often grows with the evolution time $t$. Since the bond dimension required for an accurate MPS representation typically grows exponentially with the entanglement entropy, a fixed bond dimension can lead to increasing truncation errors during long-time evolution~\cite{PAECKEL2019167998}.

\begin{figure}
    \centering
    \includegraphics[width=0.85\linewidth]{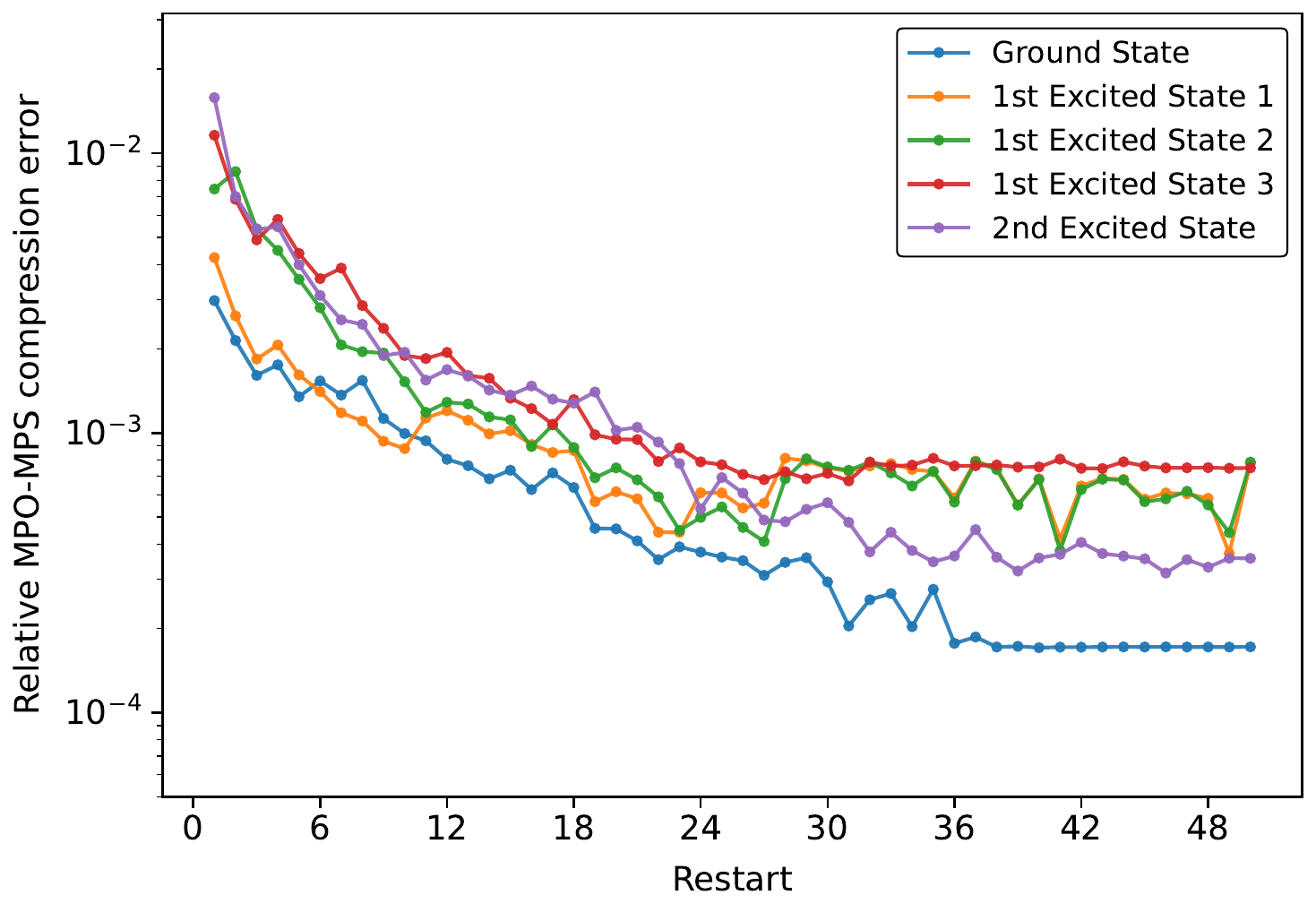}
    \caption{Relative MPO-MPS compression error for $H\ket{\phi_i}$, where $\ket{\phi_i}$ denotes a Ritz vector. The error is computed at each restart of the algorithm.}
    \label{fig:compression_error}
\end{figure}

The situation in MTBL is different. The dominant operation is not long-time evolution by $e^{-itH}$, but MPO-MPS multiplication-compressions of the form
\begin{equation}
    \ket{\psi'} = H\ket{\psi}.
\end{equation}
Let $\{\ket{e_i}\}$ denote the eigenstates of $H$, with eigenvalues $E_i$. Expanding $\ket{\psi}$ in this eigenbasis gives
\begin{equation}
    \ket{\psi} = \sum_i a_i \ket{e_i},
    \qquad
    \sum_i |a_i|^2 = 1,
\end{equation}
and therefore
\begin{equation}
    H\ket{\psi} = \sum_i a_i E_i \ket{e_i}.
\end{equation}
Thus, applying $H$ reweights the eigenstate components by their eigenvalues. Assuming that the target state is the ground state $\ket{e_0}$, i.e., $|a_0| \gg |a_i|$ for $i \ne 0$, the resulting state $H \ket{\psi}$ is even closer to the ground state since the magnitude of $E_0$ is largest. For many local Hamiltonians, low-lying eigenstates can be accurately represented by MPS with small bond dimensions due to the area law. Therefore, as the algorithm proceeds and $\ket{\psi}$ becomes closer to the ground state, one can expect that smaller and smaller bond dimensions are required to represent $H\psi$.
Equivalently, at a fixed bond dimension, the compression error in representing $H\ket{\psi}$ is not expected to grow systematically; instead, it may decrease as the iteration converges toward the ground state. To illustrate this behavior, we compute the truncation error associated with compressing $H\ket{\phi_i}$ at the beginning of each restart, where $\ket{\phi_i}$ denotes a Ritz vector. As shown in Fig.~\ref{fig:compression_error}, the compression error decreases as the Ritz vectors approach the eigenstates. Therefore, by repeatedly restarting the Lanczos procedure, we always initialize the procedure with better initial states, and the compression errors do not necessarily become larger.

\section{Benchmark on the Heisenberg Model with $J=1$, $\Delta=1$, and $h=1$}
\label{app:benchmark_heisenberg}

We also demonstrate MTBL with the nearest-neighbor exchange coupling strength $J=1$, the anisotropy parameter $\Delta = 1$, and the uniform magnetic field strength $h=1$, and the bond dimension is capped at $M=16$, and the system size is $L=16$. In such a setting, the ground state corresponds to the Tomonaga–Luttinger liquid, which is gapless in the thermodynamic limit~\cite{kono2015heisenbergchain}, and this critical regime has a dense set of low-lying excited states that makes it a numerically demanding benchmark for excited-state algorithms. 

\begin{figure}
    \centering
    \includegraphics[width=0.9\linewidth]{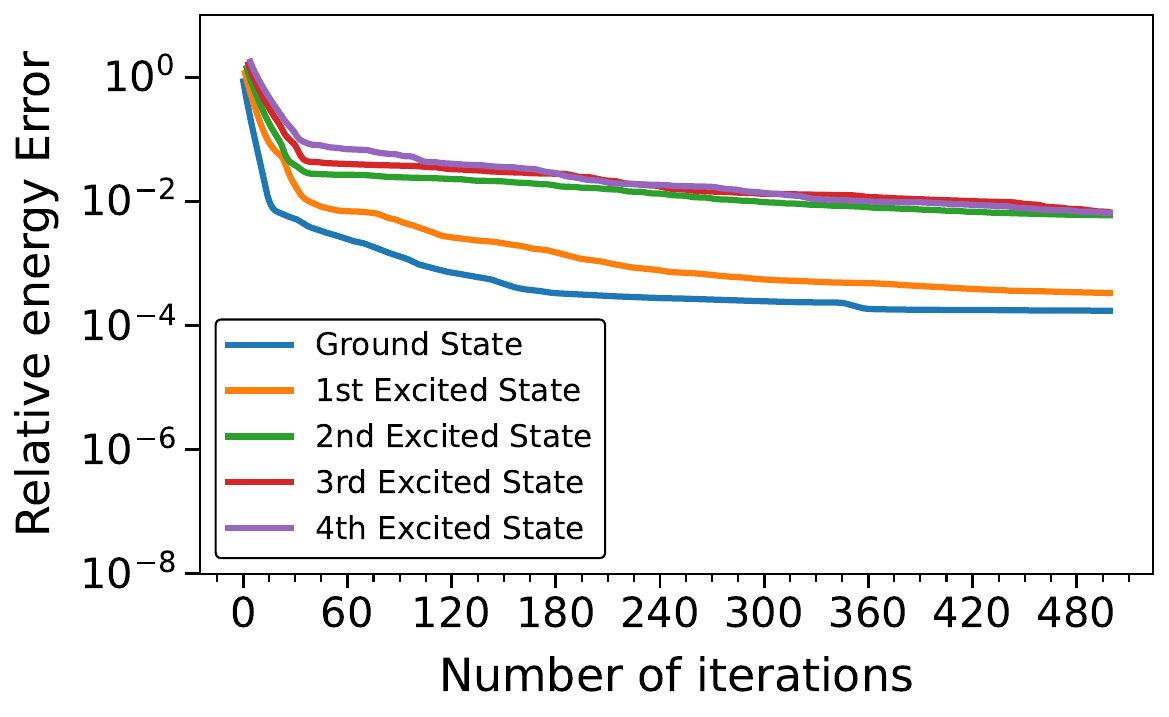}
    \caption{When using the PLM to simulate the Heisenberg model with $J=1$, $\Delta=1$, and $h=1$, the convergence stalls far away from the reference eigenvalues.}
    \label{fig:hei_plain111}
\end{figure}

Similarly, we first present the results of PLM. As shown in Fig.~\ref{fig:hei_plain111}, the convergence of all target eigenstates stalls around an error of $10^{-4}$. However, when applying MTBL, the errors of all five target states decrease to $10^{-6}$ within 180 iterations and reach their respective optimal accuracies by less than 240 steps. These results demonstrate that MTBL is both robust and broadly applicable, delivering reliable convergence across challenging excited-state problems.

\begin{figure}
    \centering
    \includegraphics[width=0.9\linewidth]{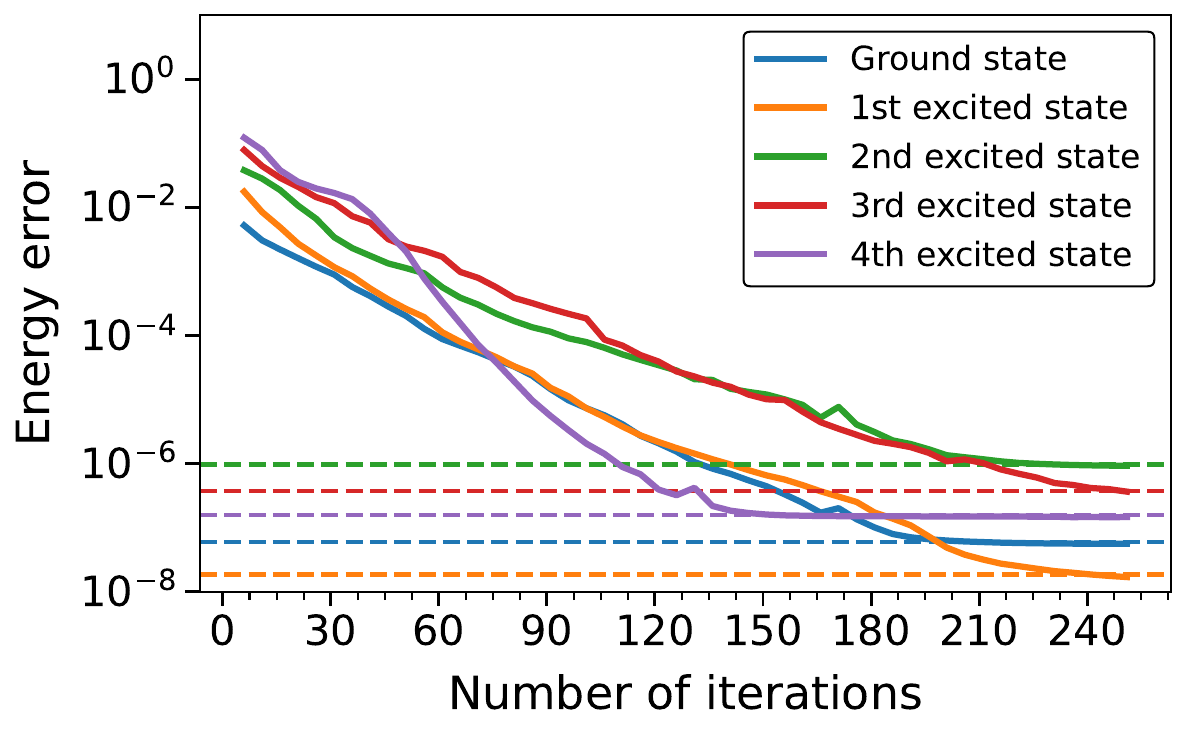}
    \caption{When using the MTBL to simulate the Heisenberg model with $L=16$, $M=16$, $J=1$, $\Delta=1$, and $h=1$, the relative energy error converges fast and reaches the optimal accuracy. For each eigenstate, the dashed line of the same color indicates the optimal accuracy achievable with the chosen bond dimension.}
    \label{fig:hei_mtrl111}
\end{figure}

\begin{figure}
    \centering
    \includegraphics[width=0.9\linewidth]{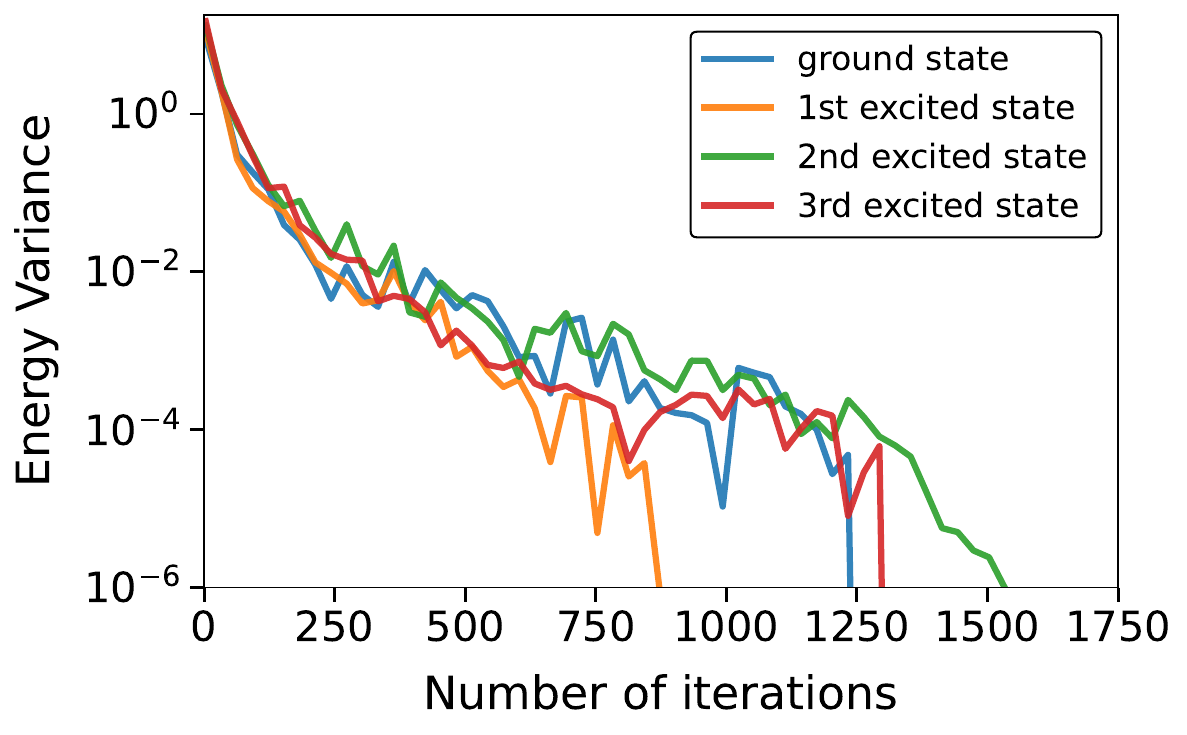}
    \caption{Track of the 4 lowest eigenstates of the spin-$1/2$ Heisenberg XXZ model in a longitudinal magnetic field with $J=1$, $\Delta=1$, and $h=1$ with $L=100$, $M=300$ in the MTBL method. We use a block size of 4 and restart the procedure every 3 iterations.}
    \label{fig:xxz100}
\end{figure}

We also extend the system size to $L=100$ and use maximum bond dimension $M=300$ to demonstrate MTBL's scalability and reliability. We use a block size of $4$, and restart the Lanczos procedure every three iterations. A target state is locked once its energy variance falls below $10^{-6}$.
As shown in Fig.~\ref{fig:xxz100}, all four target states finally reach the target energy variance.

\newpage

\bibliography{references}

@Article{dargel2012lanczos,
  author  = {Dargel, Piet E and Woellert, Anton and Honecker, Andreas and McCulloch, IP and Schollw{\"o}ck, U and Pruschke, Th},
  journal = {Phys. Rev. B},
  title   = {{Lanczos} algorithm with matrix product states for dynamical correlation functions},
  year    = {2012},
  number  = {20},
  pages   = {205119},
  volume  = {85},
  doi     = {10.1103/PhysRevB.85.205119},
}

@article{pfeifer2015symmetry,
  title   = {Symmetry-protected local minima in infinite {DMRG}},
  author  = {Pfeifer, Robert NC},
  journal = {Phys. Rev. B},
  volume  = {92},
  number  = {20},
  pages   = {205127},
  year    = {2015},
  doi     = {10.1103/PhysRevB.92.205127}
}

@article{morgan1996restarting,
  title={On restarting the Arnoldi method for large nonsymmetric eigenvalue problems},
  author={Morgan, Ronald},
  journal={Mathematics of Computation},
  volume={65},
  number={215},
  pages={1213--1230},
  year={1996},
doi = {10.1090/s0025-5718-96-00745-4 }
}

@book{golub2013matrix,
  title={Matrix computations},
  author={Golub, Gene H and Van Loan, Charles F},
  year={2013},
  publisher={JHU press},
doi = {10.56021/9781421407944 },
}

@Article{lanczos1950iteration,
  author   = {Lanczos, Cornelius},
  journal  = {J. Res. Nat. Bur. Stand.},
  title    = {An Iteration Method for the Solution of the Eigenvalue Problem of Linear Differential and Integral Operators},
  year     = {1950},
  number   = {4},
  pages    = {255--282},
  volume   = {45},
  doi      = {10.6028/jres.045.026},
}

@article{weikert1996block,
  title    = {Block {Lanczos} and many-body theory: Application to the one-particle Green’s function},
  author   = {Weikert, H-G and Meyer, H-D and Cederbaum, LS and Tarantelli, Francesco},
  journal  = {J. Chem. Phys.},
  volume   = {104},
  number   = {18},
  pages    = {7122--7138},
  year     = {1996},
}

@article{laflamme2015efficient,
  title    = {Efficient dielectric matrix calculations using the {Lanczos} algorithm for fast many-body $G_0 W_0$ implementations},
  author   = {Laflamme Janssen, Jonathan and Rousseau, Bruno and C{\^o}t{\'e}, Michel},
  journal  = {Phys. Rev. B},
  volume   = {91},
  number   = {12},
  pages    = {125120},
  year     = {2015},
  doi      = {https://doi.org/10.1103/PhysRevB.91.125120}
}

@Article{haydock1980recursive,
  author  = {Haydock, Roger},
  title   = {The recursive solution of the {S}chr\"odinger equation},
  journal = {Solid State Phys.},
  year    = {1980},
  pages   = {215--294},
  volume  = {35},
  doi     = {10.1016/S0081-1947(08)60505-6},
}

@book{Parlett1980,
  author    = {Parlett, Beresford N.},
  title     = {The Symmetric Eigenvalue Problem},
  publisher = {Society for Industrial and Applied Mathematics},
  address   = {Philadelphia},
  year      = {1998},
  series    = {Classics in Applied Mathematics},
  isbn      = {9780898714029},
  doi       = {10.1137/1.9781611971163},
}

@article{Paige1972,
  author  = {Christopher C. Paige},
  title   = {Computational variants of the {Lanczos} method for the eigenproblem},
  journal = {IMA J. Appl. Math.},
  volume  = {10},
  number  = {3},
  pages   = {373--381},
  year    = {1972},
  doi     = {10.1093/imamat/10.3.373}
}

@book{Cullum1985,
  author    = {Cullum, Jane K. and Willoughby, Ralph A.},
  title     = {{Lanczos} algorithms for large symmetric eigenvalue computations: Volume I: Theory},
  series    = {Progress in Scientific Computing},
  volume    = {3},
  publisher = {Birkhäuser Boston},
  year      = {1985},
  address   = {Boston, MA},
  doi       = {10.1007/978-1-4684-9178-4},
  isbn      = {978-1-4684-9178-4},
}

@Article{PhysRevLett.75.3537,
  author   = {\"Ostlund, Stellan and Rommer, Stefan},
  journal  = {Phys. Rev. Lett.},
  title    = {Thermodynamic Limit of Density Matrix Renormalization},
  year     = {1995},
  pages    = {3537--3540},
  volume   = {75},
  doi      = {10.1103/PhysRevLett.75.3537},
  issue    = {19},
  numpages = {0},
}

@Article{PhysRevB.55.2164,
  author   = {Rommer, Stefan and \"Ostlund, Stellan},
  journal  = {Phys. Rev. B},
  title    = {Class of ansatz wave functions for one-dimensional spin systems and their relation to the density matrix renormalization group},
  year     = {1997},
  pages    = {2164--2181},
  volume   = {55},
  doi      = {10.1103/PhysRevB.55.2164},
  issue    = {4},
  numpages = {0},
}

@Book{ma2022density,
  author = {Ma, Haibo and Schollw{\"o}ck, Ulrich and Shuai, Zhigang},
  title  = {Density Matrix Renormalization Group (DMRG)-Based Approaches in Computational Chemistry},
  year   = {2022},
  url    = {https://doi.org/10.1016/C2020-0-01314-9},
  publisher = {Elsevier},
}

@Article{schollwock2011density,
  author  = {Schollw{\"o}ck, Ulrich},
  journal = {Ann. Phys.},
  title   = {The density-matrix renormalization group in the age of matrix product states},
  year    = {2011},
  number  = {1},
  pages   = {96-192},
  volume  = {326},
  doi     = {10.1016/j.aop.2010.09.012},
}

@Article{Bridgeman_2017,
  author   = {Jacob C Bridgeman and Christopher T Chubb},
  journal  = {J. Phys. A: Math. Theor.},
  title    = {Hand-waving and interpretive dance: {A}n introductory course on tensor networks},
  year     = {2017},
  number   = {22},
  pages    = {223001},
  volume   = {50},
  doi      = {10.1088/1751-8121/aa6dc3},
}

@Article{chan2016matrix,
  author  = {Chan, Garnet Kin-Lic and Keselman, Anna and Nakatani, Naoki and Li, Zhendong and White, Steven R.},
  journal = {J. Chem. Phys.},
  title   = {Matrix product operators, matrix product states, and ab initio density matrix renormalization group algorithms},
  year    = {2016},
  number  = {1},
  pages   = {014102},
  volume  = {145},
  doi     = {10.1063/1.4955108},
}

@Article{keller2015efficient,
  author  = {Keller, Sebastian and Dolfi, Michele and Troyer, Matthias and Reiher, Markus},
  journal = {J. Chem. Phys.},
  title   = {An efficient matrix product operator representation of the quantum chemical {H}amiltonian},
  year    = {2015},
  number  = {24},
  pages   = {244118},
  volume  = {143},
  doi     = {10.1063/1.4939000},
}

@Article{PhysRevLett.69.2863,
  author   = {White, Steven R.},
  journal  = {Phys. Rev. Lett.},
  title    = {Density matrix formulation for quantum renormalization groups},
  year     = {1992},
  pages    = {2863--2866},
  volume   = {69},
  doi      = {10.1103/PhysRevLett.69.2863},
  issue    = {19},
  numpages = {0},
}

@article{white2005density,
  title    = {Density matrix renormalization group algorithms with a single center site},
  author   = {White, Steven R},
  journal  = {Phys. Rev. B},
  volume   = {72},
  number   = {18},
  pages    = {180403},
  year     = {2005},
  doi      = {10.1103/PhysRevB.72.180403}
}

@Article{PAECKEL2019167998,
  author   = {Sebastian Paeckel and Thomas K\"ohler and Andreas Swoboda and Salvatore R. Manmana and Ulrich Schollw\"ock and Claudius Hubig},
  journal  = {Annals of Physics},
  title    = {Time-evolution methods for matrix-product states},
  year     = {2019},
  pages    = {167998},
  volume   = {411},
  doi      = {10.1016/j.aop.2019.167998},
}

@article{dektor2025inexact,
  title    = {Inexact subspace projection methods for low-rank tensor eigenvalue problems},
  author   = {Dektor, Alec and DelMastro, Peter and Ye, Erika and Van Beeumen, Roel and Yang, Chao},
  journal  = {arXiv preprint arXiv:2502.19578},
  year     = {2025},
  doi      = {10.48550/arXiv.2502.19578}
}

@book{bai2000templates,
  title    = {Templates for the solution of algebraic eigenvalue problems: a practical guide},
  editor   = {Bai, Zhaojun and Demmel, James and Dongarra, Jack and Ruhe, Axel and van der Vorst, Henk},
  year     = {2000},
  publisher= {SIAM},
  doi      = {10.1137/1.9780898719581.fm}
}

@Article{ostlund1995thermodynamic,
  author   = {\"Ostlund, Stellan and Rommer, Stefan},
  journal  = {Phys. Rev. Lett.},
  title    = {Thermodynamic Limit of Density Matrix Renormalization},
  year     = {1995},
  pages    = {3537--3540},
  volume   = {75},
  doi      = {10.1103/PhysRevLett.75.3537},
  issue    = {19},
  numpages = {0},
}

@Article{rommer1997class,
  author   = {Rommer, Stefan and \"Ostlund, Stellan},
  journal  = {Phys. Rev. B},
  title    = {Class of ansatz wave functions for one-dimensional spin systems and their relation to the density matrix renormalization group},
  year     = {1997},
  pages    = {2164--2181},
  volume   = {55},
  doi      = {10.1103/PhysRevB.55.2164},
  issue    = {4},
  numpages = {0},
}

@Article{ren2020general,
  author  = {Ren, Jiajun and Li, Weitang and Jiang, Tong and Shuai, Zhigang},
  journal = {J. Chem. Phys.},
  title   = {A general automatic method for optimal construction of matrix product operators using bipartite graph theory},
  year    = {2020},
  number  = {8},
  pages   = {084118},
  volume  = {153},
  doi     = {10.1063/5.0018149},
}

@article{cakir2025optimal,
  author       = {Hazar Çakır and Richard M. Milbradt and Christian B. Mendl},
  title        = {Optimal Symbolic Construction of Matrix Product Operators and Tree Tensor Network Operators},
  journal      = {arXiv preprint arXiv:2502.18630},
  year         = {2025},
  doi          = {10.48550/arXiv.2502.18630}
}

@Article{hauschild2018efficient,
  title   = {Efficient numerical simulations with tensor networks: {T}ensor {N}etwork {P}ython ({TeNPy})},
  author  = {Hauschild, Johannes and Pollmann, Frank},
  journal = {SciPost Phys. Lect. Notes},
  pages   = {5},
  year    = {2018},
  doi     = {10.21468/SciPostPhysLectNotes.5},
}

@Article{Stoudenmire_2010,
  author   = {E M Stoudenmire and Steven R White},
  journal  = {New J. Phys.},
  title    = {Minimally entangled typical thermal state algorithms},
  year     = {2010},
  number   = {5},
  pages    = {055026},
  volume   = {12},
  doi      = {10.1088/1367-2630/12/5/055026},
}

@article{ma2024approximate,
  doi      = {10.22331/q-2024-12-27-1580},
  title    = {Approximate {C}ontraction of {A}rbitrary {T}ensor {N}etworks with a {F}lexible and {E}fficient {D}ensity {M}atrix {A}lgorithm},
  author   = {Ma, Linjian and Fishman, Matthew and Stoudenmire, Edwin Miles and Solomonik, Edgar},
  journal  = {{Quantum}},
  issn     = {2521-327X},
  volume   = {8},
  pages    = {1580},
  year     = {2024},
}

@article{Heisenberg1928,
  author  = {W. Heisenberg},
  title   = {Zur Theorie des Ferromagnetismus},
  journal = {Zeitschrift f{\"u}r Physik},
  volume  = {49},
  pages   = {619--636},
  year    = {1928},
  doi     = {10.1007/BF01328601}
}

@book{cullum2002lanczos,
  author    = {Cullum, Jane K. and Willoughby, Ralph A.},
  title     = {{Lanczos} Algorithms for Large Symmetric Eigenvalue Computations},
  publisher = {SIAM},
  year      = {2002},
  doi       = {10.1137/1.9780898719192}
}

@book{martin2004electronic,
  author    = {Martin, Richard M.},
  title     = {Electronic Structure: Basic Theory and Practical Methods},
  publisher = {Cambridge University Press},
  year      = {2004},
  doi       = {10.1017/CBO9780511805769}
}

@article{Hubbard1963,
  author  = {Hubbard, J.},
  title   = {Electron Correlations in Narrow Energy Bands},
  journal = {Proc. R. Soc. Lond. A},
  volume  = {276},
  number  = {1365},
  pages   = {238--257},
  year    = {1963},
  doi     = {10.1098/rspa.1963.0204},
}

@article{Georges1996,
  author  = {Georges, A. and Kotliar, G. and Krauth, W. and Rozenberg, M. J.},
  title   = {Dynamical mean‐field theory of strongly correlated fermion systems and the limit of infinite dimensions},
  journal = {Rev. Mod. Phys.},
  volume  = {68},
  pages   = {13--125},
  year    = {1996},
  doi     = {10.1103/RevModPhys.68.13},
}

@article{Jordens2008,
  author  = {J\"ordens, R. and Strohmaier, N. and G\"unter, K. and Moritz, H. and Esslinger, T.},
  title   = {A Mott insulator of fermionic atoms in an optical lattice},
  journal = {Nature},
  volume  = {455},
  pages   = {204--207},
  year    = {2008},
  doi     = {10.1038/nature07244},
}

@article{Hofrichter2016,
  author  = {Hofrichter, C. and Riegger, L. and Scazza, F. and H\"ofer, M. and Rio Fernandes, D. and Bloch, I. and F\"olling, S.},
  title   = {Direct Probing of the Mott Crossover in the SU(N) Fermi-Hubbard Model},
  journal = {Phys. Rev. X},
  volume  = {6},
  issue   = {2},
  pages   = {021030},
  year    = {2016},
  doi     = {10.1103/PhysRevX.6.021030},
}

@article{Anderson1987,
  author  = {Anderson, P.~W.},
  title   = {The Resonating Valence Bond State in {La}$_2${CuO}$_4$ and Superconductivity},
  journal = {Science},
  volume  = {235},
  number  = {4793},
  pages   = {1196--1198},
  year    = {1987},
  doi     = {10.1126/science.235.4793.1196},
}

@article{Scalapino1986,
  author  = {Scalapino, D. J. and {Loh Jr.}, E. and Hirsch, J. E.},
  journal = {Phys. Rev. B},
  title   = {{d}-wave pairing near a spin-density-wave instability},
  year    = {1986},
  number  = {11},
  pages   = {8190-8192},
  volume  = {34},
  doi     = {10.1103/PhysRevB.34.8190},
}

@article{Dagotto1994,
  author  = {Dagotto, E.},
  title   = {Correlated Electrons in High-Temperature Superconductors},
  journal = {Rev. Mod. Phys.},
  volume  = {66},
  number  = {3},
  pages   = {763--840},
  year    = {1994},
  doi     = {10.1103/RevModPhys.66.763},
}

@article{YangYang1966I,
  author  = {Yang, C.~N. and Yang, C.~P.},
  title   = {One-Dimensional Chain of Anisotropic Spin-Spin Interactions. I. Proof of Bethe's Hypothesis for the Ground State in a Finite System},
  journal = {Phys. Rev.},
  volume  = {150},
  pages   = {321--327},
  year    = {1966},
  doi     = {10.1103/PhysRev.150.321},
}

@article{YangYang1966II,
  author  = {Yang, C.~N. and Yang, C.~P.},
  title   = {One-Dimensional Chain of Anisotropic Spin-Spin Interactions. II. Properties of the Ground-State Energy per Lattice Site for an Infinite System},
  journal = {Phys. Rev.},
  volume  = {150},
  pages   = {327--339},
  year    = {1966},
  doi     = {10.1103/PhysRev.150.327},
}

@article{Alcaraz1987,
  author  = {Alcaraz, F.~C. and Barber, M.~N. and Batchelor, M.~T. and Baxter, R.~J. and Quispel, G.~R.~W.},
  title   = {Surface Exponents of the Quantum {XXZ}, Ashkin–Teller and Potts Models},
  journal = {J. Phys. A: Math. Gen.},
  volume  = {20},
  pages   = {6397--6409},
  year    = {1987},
  doi     = {10.1088/0305-4470/20/18/038},
}

@article{Bertini2016,
  author  = {Bertini, Bruno and Collura, Mario and De Nardis, Jacopo and Fagotti, Maurizio},
  title   = {Transport in Out-of-Equilibrium {XXZ} Chains: Exact Profiles of Charges and Currents},
  journal = {Phys. Rev. Lett.},
  volume  = {117},
  pages   = {207201},
  year    = {2016},
  doi     = {10.1103/PhysRevLett.117.207201},
}

@Book{avella2013stronglycorrelated,
  editor    = {Avella, Adolfo and Mancini, Ferdinando},
  title     = {Strongly Correlated Systems: Numerical Methods},
  publisher = {Springer},
  address   = {Berlin, Heidelberg},
  year      = {2013},
  edition   = {1},
  series    = {Springer Series in Solid-State Sciences},
  volume    = {176},
  isbn      = {978-3-642-35106-8},
  doi       = {10.1007/978-3-642-35106-8},
}

@article{GaglianoBalseiro1987,
  author  = {Gagliano, E.~R. and Balseiro, C.~A.},
  title   = {Dynamical properties of quantum many-body systems at zero temperature},
  journal = {Phys. Rev. Lett.},
  volume  = {59},
  pages   = {2999--3002},
  year    = {1987},
  doi     = {10.1103/PhysRevLett.59.2999},
}

@article{Verstraete2006PEPS,
  author  = {Verstraete, F. and Wolf, M. M. and Pérez-García, D. and Cirac, J. I.},
  title   = {Criticality, the area law, and the computational power of projected entangled pair states},
  journal = {Phys. Rev. Lett.},
  volume  = {96},
  issue   = {22},
  pages   = {220601},
  year    = {2006},
  doi     = {10.1103/PhysRevLett.96.220601},
}

@article{Shi2006TTNS,
  author  = {Shi, Y.-Y. and Duan, L.-M. and Vidal, G.},
  title   = {Classical simulation of quantum many-body systems with a tree tensor network},
  journal = {Phys. Rev. A},
  volume  = {74},
  pages   = {022320},
  year    = {2006},
  doi     = {10.1103/PhysRevA.74.022320},
}

@article{Zaletel2020isoTNS,
  author  = {Zaletel, Michael P. and Pollmann, Frank},
  title   = {Isometric Tensor Network States in Two Dimensions},
  journal = {Phys. Rev. Lett.},
  volume  = {124},
  pages   = {037201},
  year    = {2020},
  doi     = {10.1103/PhysRevLett.124.037201},
}

@book{LehoucqSorensenYang1998,
  author    = {Lehoucq, Richard B. and Sorensen, Danny C. and Yang, Chao},
  title     = {{ARPACK Users’ Guide: Solution of Large-Scale Eigenvalue Problems with Implicitly Restarted Arnoldi Methods}},
  series    = {Software, Environments \& Tools},
  volume    = {6},
  publisher = {Society for Industrial and Applied Mathematics},
  year      = {1998},
  doi       = {10.1137/1.9780898719628},
  isbn      = {978-0-89871-407-4},
}

@article{SandvikSSE1997,
  author  = {Sandvik, A.~W.},
  title   = {Finite-size scaling of the ground-state parameters of the two-dimensional {Heisenberg} model},
  journal = {Phys. Rev. B},
  volume  = {56},
  pages   = {11678--11690},
  year    = {1997},
  doi     = {10.1103/PhysRevB.56.11678},
}

@Article{banuls2023routemap,
  author  = {Ba{\~n}uls, Mari Carmen},
  journal = {Annu. Rev. Condens. Matter Phys.},
  title   = {Tensor Network Algorithms: A Route Map},
  year    = {2023},
  number  = {1},
  pages   = {173-191},
  volume  = {14},
  doi     = {10.1146/annurev-conmatphys-040721-022705},
}

@article{SugiuraShimizu2013,
  author  = {Sugiura, Sho and Shimizu, Akira},
  title   = {Canonical Thermal Pure Quantum State},
  journal = {Phys. Rev. Lett.},
  volume  = {111},
  number  = {1},
  pages   = {010401},
  year    = {2013},
  doi     = {10.1103/PhysRevLett.111.010401},
}

@article{JaklicPrelovsek2000,
  author  = {Jaklič, J. and Prelovšek, P.},
  title   = {Finite-temperature properties of doped antiferromagnets},
  journal = {Adv. Phys.},
  volume  = {49},
  number  = {1},
  pages   = {1--92},
  year    = {2000},
  doi     = {10.1080/000187300243381},
}

@article{laguta2023low,
  title={Low-temperature ground state structure of {PbTiO$_3$}},
  author={Laguta, V and Zagorodniy, Yu O and Kuzian, RO and Kondakova, IV and Chlan, V and {\v{R}}ezn{\'\i}{\v{c}}ek, R and {\v{S}}t{\v{e}}p{\'a}nkov{\'a}, H and Bohdanov, D and Hlinka, J and Ramesh, R},
  journal={Phys. Rev. B},
  volume={107},
  number={10},
  pages={104107},
  year={2023},
  publisher={APS},
  doi={10.1103/PhysRevB.107.104107}
}

@article{SunMotta2021,
  author  = {Sun, Shi‐Ning and Motta, Mario and Tazhigulov, Ruslan N. and Tan, Adrian T. K. and Chan, Garnet Kin‐Lic and Minnich, Austin J.},
  title   = {Quantum Computation of Finite-Temperature Static and Dynamical Properties of Spin Systems Using Quantum Imaginary Time Evolution},
  journal = {PRX Quantum},
  volume  = {2},
  pages   = {010317},
  year    = {2021},
  doi     = {10.1103/PRXQuantum.2.010317},
}

@article{WeinbergBukov2019,
  author  = {Weinberg, Phillip and Bukov, Marin},
  title   = {QuSpin: a Python package for dynamics and exact diagonalisation of quantum many body systems. Part II: bosons, fermions and higher spins},
  journal = {SciPost Phys.},
  volume  = {7},
  pages   = {020},
  year    = {2019},
  doi     = {10.21468/SciPostPhys.7.2.020}
}

@article{LauchliSudanMoessner2019,
  author  = {Läuchli, Andreas M. and Sudan, Justin and Moessner, Roderich},
  title   = {The $S=1/2$ kagome {Heisenberg} antiferromagnet revisited},
  journal = {Phys. Rev. B},
  volume  = {100},
  pages   = {155142},
  year    = {2019},
  doi     = {10.1103/PhysRevB.100.155142}
}

@article{IskakovDanilov2018,
  author  = {Iskakov, Sergei and Danilov, Michael},
  journal = {Comput. Phys. Commun.},
  title   = {Exact diagonalization library for quantum electron models},
  year    = {2018},
  pages   = {128-139},
  volume  = {225},
  doi     = {10.1016/j.cpc.2017.12.016},
}

@article{jiang2021chebyshev,
  title     = {Chebyshev matrix product states with canonical orthogonalization for spectral functions of many-body systems},
  author    = {Jiang, Tong and Ren, Jiajun and Shuai, Zhigang},
  journal   = {J. Phys. Chem. Lett.},
  volume    = {12},
  number    = {38},
  pages     = {9344--9352},
  year      = {2021},
  doi       = {10.1021/acs.jpclett.1c02688}
}

@article{wu2000thick,
  author  = {K. K. Wu and H. Simon},
  title   = {Thick-Restart {Lanczos} Method for Large Symmetric Eigenvalue Problems},
  journal = {SIAM J. Matrix Anal. Appl.},
  volume  = {22},
  number  = {2},
  pages   = {602--616},
  year    = {2000},
  doi     = {10.1137/S0895479898334605},
}

@article{hackett2025blocklanczos,
  author  = {D. C. Hackett and M. L. Wagman},
  title   = {Block {Lanczos} algorithm for lattice QCD spectroscopy and matrix elements},
  journal = {Phys. Rev. D},
  volume  = {112},
  pages   = {014514},
  year    = {2025},
  doi     = {10.1103/fp74-q35q},
}

@incollection{golub1977block,
  author    = {G. H. Golub and R. Underwood},
  title     = {The Block {Lanczos} Method for Computing Eigenvalues},
  editor    = {J. R. Rice},
  booktitle = {Mathematical Software},
  publisher = {Academic Press},
  pages     = {361--377},
  year      = {1977},
  isbn      = {978-0-12-587260-7},
  doi       = {10.1016/B978-0-12-587260-7.50018-2}
}

@article{saad1980rates,
  author  = {Y. Saad},
  title   = {On the Rates of Convergence of the {Lanczos} and the Block-{Lanczos} Methods},
  journal = {SIAM J. Numer. Anal.},
  volume  = {17},
  number  = {5},
  pages   = {687--706},
  year    = {1980},
  doi     = {10.1137/0717059}
}

@Article{wu1999thickrestart,
  author  = {Wu, Kesheng and Canning, Andrew and Simon, Horst D. and Wang, Lin-Wang},
  journal = {J. Comput. Phys.},
  title   = {Thick-Restart {Lanczos} Method for Electronic Structure Calculations},
  year    = {1999},
  number  = {1},
  pages   = {156-173},
  volume  = {154},
  doi     = {10.1006/jcph.1999.6306},
}

@article{dagotto1994correlated,
  author  = {E. Dagotto},
  title   = {Correlated electrons in high-temperature superconductors},
  journal = {Rev. Mod. Phys.},
  volume  = {66},
  number  = {3},
  pages   = {763--840},
  year    = {1994},
  doi     = {10.1103/RevModPhys.66.763},
}

@article{sandvik2010computational,
  author  = {A. W. Sandvik},
  title   = {Computational Studies of Quantum Spin Systems},
  journal = {AIP Conference Proceedings},
  volume  = {1297},
  number  = {1},
  pages   = {135--338},
  year    = {2010},
  month   = {11},
  doi     = {10.1063/1.3518900}
}

@article{gleis2022projector,
  author  = {Andreas Gleis and Jheng-Wei Li and Jan von Delft},
  title   = {Projector formalism for kept and discarded spaces of matrix product states},
  journal = {Phys. Rev. B},
  volume  = {106},
  pages   = {195138},
  year    = {2022},
  doi     = {10.1103/PhysRevB.106.195138},
}

@article{hubig2018errorestimates,
  author  = {C. Hubig and J. Haegeman and U. Schollw\"ock},
  title   = {Error estimates for extrapolations with matrix-product states},
  journal = {Phys. Rev. B},
  volume  = {97},
  number  = {4},
  pages   = {045125},
  year    = {2018},
  doi     = {10.1103/PhysRevB.97.045125},
}

@article{gleis2023controlled,
  author  = {Andreas Gleis and Jheng-Wei Li and Jan von Delft},
  title   = {Controlled Bond Expansion for Density Matrix Renormalization Group Ground State Search at Single-Site Costs},
  journal = {Phys. Rev. Lett.},
  volume  = {130},
  number  = {24},
  pages   = {246402},
  year    = {2023},
  doi     = {10.1103/PhysRevLett.130.246402},
}

@Article{pytenet,
  author  = {Mendl, Christian B.},
  title   = {{P}y{T}e{N}et: A concise {P}ython implementation of quantum tensor network algorithms},
  journal = {J. Open Source Softw.},
  year = {2018},
  volume  = {3},
  number  = {30},
  pages   = {948},
  doi     = {10.21105/joss.00948},
}

@article{huang2018generalized,
  author  = {Rui-Zhen Huang and Hai-Jun Liao and Zhi-Yuan Liu and Hai-Dong Xie and Zhi-Yuan Xie and Hui-Hai Zhao and Jing Chen and Tao Xiang},
  title   = {Generalized {Lanczos} method for systematic optimization of tensor network states},
  journal = {Chinese Physics B},
  volume  = {27},
  number  = {7},
  pages   = {070501},
  year    = {2018},
  doi     = {10.1088/1674-1056/27/7/070501},
}

@misc{li2024boundstates,
  author        = {P. Li and Y. Shen and M. Qin and K. Jiang and J. Hu and T. Xiang},
  title         = {Bound states in doped charge transfer insulators},
  year          = {2024},
  eprint        = {2408.00576},
  archivePrefix = {arXiv},
  doi           = {10.48550/arXiv.2408.00576},
}

@misc{paeckel2023bandlanczos,
  author        = {S. Paeckel and T. K\"ohler and S. R. Manmana and B. Lenz},
  title         = {Matrix-product-state-based band-{Lanczos} solver for quantum cluster approaches},
  year          = {2023},
  eprint        = {2310.10799},
  archivePrefix = {arXiv},
  doi           = {10.48550/arXiv.2310.10799},
}

@article{camano2025src,
  author        = {Chris Cam{\~a}no and Ethan N. Epperly and Joel A. Tropp},
  title         = {Successive randomized compression: A randomized algorithm for the compressed {MPO}–{MPS} product},
  journal       = {arXiv preprint arXiv:2504.06475},
  year          = {2025},
  archivePrefix = {arXiv},
  eprint        = {2504.06475},
}

@article{kono2015heisenbergchain,
  author  = {Y. Kono and T. Sakakibara and C. P. Aoyama and C. Hotta and M. M. Turnbull and C. P. Landee and Y. Takano},
  title   = {Field-Induced Quantum Criticality and Universal Temperature Dependence of the Magnetization of a Spin-1/2 {Heisenberg} Chain},
  journal = {Phys. Rev. Lett.},
  volume  = {114},
  number  = {3},
  pages   = {037202},
  year    = {2015},
  doi     = {10.1103/PhysRevLett.114.037202},
}

@Article{rano2025inexactlanczos,
  author  = {Rano, Madhumita and Larsson, Henrik R.},
  journal = {J. Chem. Phys.},
  title   = {Computing excited eigenstates using inexact {Lanczos} methods and tree tensor network states},
  year    = {2025},
  number  = {16},
  pages   = {164110},
  volume  = {163},
  doi     = {10.1063/5.0301263},
}

@Article{baiardi2019optimization,
  author  = {Baiardi, Alberto and Stein, Christopher J. and Barone, Vincenzo and Reiher, Markus},
  journal = {J. Chem. Phys.},
  title   = {Optimization of highly excited matrix product states with an application to vibrational spectroscopy},
  year    = {2019},
  number  = {9},
  pages   = {094113},
  volume  = {150},
  doi     = {10.1063/1.5068747},
}

@Article{dorando2007targeted,
  author  = {Dorando, Jonathan J. and Hachmann, Johannes and Chan, Garnet Kin-Lic},
  journal = {J. Chem. Phys.},
  title   = {Targeted excited state algorithms},
  year    = {2007},
  number  = {8},
  pages   = {084109},
  volume  = {127},
  doi     = {10.1063/1.2768360},
}

@Article{tran2019tracking,
  author  = {Tran, Lan Nguyen and Shea, Jacqueline A. R. and Neuscamman, Eric},
  journal = {J. Chem. Theory Comput.},
  title   = {Tracking Excited States in Wave Function Optimization Using Density Matrices and Variational Principles},
  year    = {2019},
  number  = {9},
  pages   = {4790-4803},
  volume  = {15},
  doi     = {10.1021/acs.jctc.9b00351},
}

@Book{wilkinson1965algebraic,
  author    = {Wilkinson, James H.},
  title     = {The Algebraic Eigenvalue Problem},
  publisher = {Oxford University Press},
  address   = {Oxford},
  year      = {1965},
  series    = {Monographs on Numerical Analysis},
  isbn      = {0198534035},
}

@Article{ruhe1984rationalkrylov,
  author  = {Ruhe, Axel},
  journal = {Linear Algebra Appl.},
  title   = {Rational {Krylov} sequence methods for eigenvalue computation},
  year    = {1984},
  pages   = {391-405},
  volume  = {58},
  doi     = {10.1016/0024-3795(84)90221-0},
}

@Book{saad2011largEeig,
  author    = {Saad, Yousef},
  title     = {Numerical Methods for Large Eigenvalue Problems},
  publisher = {Society for Industrial and Applied Mathematics},
  address   = {Philadelphia},
  year      = {2011},
  edition   = {Revised},
  series    = {Classics in Applied Mathematics},
  isbn      = {978-1-61197-072-2},
  doi       = {10.1137/1.9781611970739},
}

@Article{pietracaprina2018shiftinvert,
  author  = {Pietracaprina, Francesca and Mac{\'e}, Nicolas and Luitz, David J. and Alet, Fabien},
  journal = {SciPost Phys.},
  title   = {Shift-invert diagonalization of large many-body localizing spin chains},
  year    = {2018},
  pages   = {045},
  volume  = {5},
  doi     = {10.21468/SciPostPhys.5.5.045},
}

@Article{yu2017finding,
  author  = {Yu, Xiongjie and Pekker, David and Clark, Bryan K.},
  journal = {Phys. Rev. Lett.},
  title   = {Finding Matrix Product State Representations of Highly Excited Eigenstates of Many-Body Localized Hamiltonians},
  year    = {2017},
  number  = {1},
  pages   = {017201},
  volume  = {118},
  doi     = {10.1103/PhysRevLett.118.017201},
}

@Article{rakhuba2016vibtt,
  author  = {Rakhuba, Maxim and Oseledets, Ivan},
  journal = {J. Chem. Phys.},
  title   = {Calculating vibrational spectra of molecules using tensor train decomposition},
  year    = {2016},
  number  = {12},
  pages   = {124101},
  volume  = {145},
  doi     = {10.1063/1.4962420},
}

@Article{pomata2023seeking,
  author  = {Pomata, Nicholas and Ganeshan, Sriram and Wei, Tzu-Chieh},
  journal = {Phys. Rev. B},
  title   = {Seeking a many-body mobility edge with matrix product states in a quasiperiodic model},
  year    = {2023},
  number  = {9},
  pages   = {094201},
  volume  = {108},
  doi     = {10.1103/PhysRevB.108.094201},
}

@Article{szenes2025qcmaquis4,
  author  = {Szenes, Kalman and Glaser, Nina and Erakovi{\'c}, Mihael and Barandun, Valentin and M{\"o}rchen, Maximilian and Feldmann, Robin and Battaglia, Stefano and Baiardi, Alberto and Reiher, Markus},
  journal = {J. Phys. Chem. A},
  title   = {{QCMaquis} 4.0: Multipurpose Electronic, Vibrational, and Vibronic Structure and Dynamics Calculations with the Density Matrix Renormalization Group},
  year    = {2025},
  number  = {32},
  pages   = {7549-7574},
  volume  = {129},
  doi     = {10.1021/acs.jpca.5c02970},
}

@Article{larsson2019ttns,
  author  = {Larsson, Henrik R.},
  journal = {J. Chem. Phys.},
  title   = {Computing vibrational eigenstates with tree tensor network states ({TTNS})},
  year    = {2019},
  number  = {20},
  pages   = {204102},
  volume  = {151},
  doi     = {10.1063/1.5130390},
}

@article{Bhattacharya2023,
  author  = {Bhattacharya, Aranya and Nandy, Pratik and Nath, Pingal Pratyush and Sahu, Himanshu},
  title   = {On {Krylov} complexity in open systems: an approach via bi-{Lanczos} algorithm},
  journal = {J. High Energy Phys.},
  year    = {2023},
  month   = {Dec},
  volume  = {2023},
  number   = {66},
  doi     = {10.1007/JHEP12(2023)066},
}

@article{larsson2026accurate,
  title={Accurate, full-dimensional computations of thousands of complex vibrational eigenstates with tree tensor network states},
  author={Larsson, Henrik R and D{\'e}, Brieuc Le and Gamboni, Gino E},
  journal={arXiv preprint arXiv:2605.00998},
  year={2026},
  doi={10.48550/arXiv.2605.00998}
}

@article{baker2024bundled,
  title={Bundled matrix product states represent low-energy excitations faithfully},
  author={Baker, Thomas E and Seif, Negar},
  journal={Journal of Physics A: Mathematical and Theoretical},
  volume={57},
  number={39},
  pages={395306},
  year={2024},
  doi={10.1088/1751-8121/ad770f}
}

@article{baker2024direct,
  title={Direct solution of multiple excitations in a matrix product state with block Lanczos},
  author={Baker, Thomas E and Foley, Alexandre and S{\'e}n{\'e}chal, David},
  journal={The European Physical Journal B},
  volume={97},
  number={6},
  pages={72},
  year={2024},
  doi={10.1140/epjb/s10051-024-00702-7}
}

@article{baker2024qc,
  title = {Block Lanczos method for excited states on a quantum computer},
  author = {Baker, Thomas E.},
  journal = {Phys. Rev. A},
  volume = {110},
  issue = {1},
  pages = {012420},
  numpages = {11},
  year = {2024},
  month = {Jul},
  publisher = {American Physical Society},
  doi = {10.1103/PhysRevA.110.012420}
}

@article{lixuan2024mtu,
  title = {Accurate determination of low-energy eigenspectra with multitarget matrix product states},
  author = {Li, Xuan and Zhou, Zongsheng and Xu, Guanglei and Chi, Runze and Guo, Yibin and Liu, Tong and Liao, Haijun and Xiang, Tao},
  journal = {Phys. Rev. B},
  volume = {109},
  issue = {4},
  pages = {045115},
  numpages = {12},
  year = {2024},
  month = {Jan},
  doi = {10.1103/PhysRevB.109.045115}
}

@article{prgjndfm,
  title = {Excited states from local effective Hamiltonians of matrix product states and their entanglement spectrum transition},
  author = {Cocchiarella, Denise and Yang, Mingru and Zhang, Yueshui and Ba\~nuls, Mari Carmen and Tu, Hong-Hao and Liu, Yuhan},
  journal = {Phys. Rev. B},
  volume = {113},
  issue = {20},
  pages = {205145},
  numpages = {12},
  year = {2026},
  month = {May},
  doi = {10.1103/prgj-ndfm},
}

@misc{datas,
  author       = {Wang, Yu and Yang, Zhangyu and Wu, Xingyao and Mendl, Christian B.},
  title        = {Data repository for ``{A High-Accuracy Lanczos Method in the Matrix Product State Representation}''},
  year         = {2026},
  publisher    = {Zenodo},
  doi          = {10.5281/zenodo.20625898},
  url          = {https://doi.org/10.5281/zenodo.20625898}
}
\end{document}